        \documentclass[useAMS,usenatbib]{mn2e}
        \usepackage{graphicx}

\def\oversim#1#2{\lower0.5pt\vbox{\baselineskip0pt \lineskip-0.5pt
     \ialign{$\mathsurround0pt #1\hfil##\hfil$\crcr#2\crcr\sim\crcr}}}

\title[Obscured AGB Variables in the LMC]
{Obscured Asymptotic Giant Branch Variables in the
Large Magellanic Cloud and the Period-Luminosity Relation\thanks{This
paper is based on observations made at the South African
Astronomical Observatory.}}

\author[P.A. Whitelock et al.]
{Patricia A. Whitelock$^1$\thanks{e-mail: \tt paw@saao.ac.za}, 
    Michael W. Feast$^2$, 
       Jacco Th. van Loon$^3$ and  \newauthor Albert A.  Zijlstra$^4$\\
      $^1$ South African Astronomical Observatory, P.O.Box 9, 7935
           Observatory, South Africa.\\
       $^2$ Astronomy Department, University of Cape Town, 7701 Rondebosch,
           South Africa.\\
	$^3$ Astrophysics Group, School of Chemistry \& Physics, Keele
	University, Staffordshire ST5 5BG, UK. \\
	$^4$ Department of Physics, UMIST, PO Box 88, Manchester M60 1QD,
	UK.\\
}
\date{Received date; accepted date}

\pubyear{2003}
\begin{document}
\maketitle

\begin{abstract} The characteristics of oxygen-rich and 
carbon-rich, large amplitude ($\Delta K>0.4$ mag), asymptotic giant branch
variables in the Large Magellanic Clouds are discussed, with an emphasis on
those obscured by dust. Near-infrared photometry, obtained over about 8
years, is combined with published mid-infrared observations from IRAS and
ISO to determine bolometric magnitudes for 42 stars. Pulsation periods of
the O-rich stars are in the range $116<P<1393$ days, while those for C-rich
stars have $298<P<939$ days. In addition to the regular pulsations, one
O-rich star and four C-rich stars show large amplitude, $\Delta K> 0.6$ mag,
secular or very long period variations which may be associated with changes
in their mass-loss rates. We discuss and compare various methods of
determining the bolometric magnitudes and show, perhaps surprisingly, that
most of the very long period stars seem to follow an extrapolation of the 
period-luminosity relation determined for stars with shorter periods -
although the details do depend on how the bolometric magnitudes are
calculated.

Three stars with thin shells, which are clearly more luminous than the
obscured AGB stars, are undergoing hot bottom burning, while other stars
with similar luminosities have yet to be investigated in sufficient detail
to determine their status in this regard. We suggest that an apparent
change in slope of the period luminosity relation around 400-420 days is
caused by variables with luminosities brighter than the predictions of the
core-mass luminosity relation, due to excess flux from hot bottom burning.
 \end{abstract}


%
%
\begin{table}
\begin{center}
\caption[]{\label{table.JHKL} Near-infrared photometry for the LMC stars}
\begin{tabular}{rrrrr}
\hline
  JD& $J$ & $H$ & $K$ & $L$ \\ 
--2440000& \multicolumn{4}{c}{(mag)}\\
\hline
\multicolumn{5}{l}{{\it O-rich stars}}\\
\multicolumn{5}{l}{HV\,12070  (WOH\,SG\,515)}  \\
  4287.4 & 10.6  &  9.6  &  9.13 &    *  \\ 
  4551.6 & 10.6  &  9.7  &  9.27 &    *  \\ 
  4889.6 & 10.58 &  9.65 &  9.15 &       \\
  4932.5 & 10.99 & 10.07 &  9.46 &    *  \\
  5011.5 & 10.76 &  9.91 &  9.37 &    *  \\
  5259.6 & 10.00 &  9.04 &  8.68 &    *  \\
  5331.5 &  9.89 &  8.88 &  8.52 &       \\
  5648.5 & 10.51 &  9.52 &  9.11 &    *  \\
  6034.5 & 10.09 &  9.07 &  8.67 &    *  \\
  6359.5 & 10.65 &  9.81 &  9.36 &    *  \\
  6426.5 & 10.76 &  9.85 &  9.34 &    *  \\
  6442.4 & 10.63 &  9.73 &  9.29 &    *  \\
  6491.4 & 10.16 &  9.26 &  8.87 &    *  \\
  7160.6 & 10.00 &  9.00 &  8.65 &       \\
 10035.5 & 10.73 &  9.88 &  9.37 &  8.76 \\
 10121.5 & 10.73 &  9.81 &  9.35 &  8.79 \\
 10201.3 & 10.26 &  9.36 &  9.04 &  8.48 \\
 10324.7 &  9.93 &  8.89 &  8.57 &  8.16 \\
 10447.6 & 10.28 &  9.28 &  8.86 &  8.55 \\
 10473.5 & 10.48 &  9.48 &  9.00 &  8.62 \\
 10917.4 &  9.91 &  8.82 &  8.48 &  7.96 \\
\multicolumn{5}{l}{HV\,2446 (WOH\,G274)}\\
  4890.5 & 10.09 &  9.04 &  8.61 &       \\
  4927.5 & 10.31 &  9.26 &  8.79 &    *  \\
  5036.4 & 10.91 & 10.00 &  9.44 &    *  \\
  5679.5 & 10.48 &  9.51 &  9.03 &    *  \\
  6034.5 & 10.08 &  9.01 &  8.64 &    *  \\
  6360.4 & 10.70 &  9.82 &  9.36 &    *  \\
  6429.4 & 10.95 & 10.07 &  9.63 &    *  \\
  6491.3 & 10.44 &  9.50 &  9.20 &    *  \\
  6794.4 & 10.66 &  9.68 &  9.18 &    *  \\
  9961.7 & 10.56 &  9.65 &  9.19 &  8.60 \\
 10029.4 & 10.78 &  9.82 &  9.34 &  8.5\,\,\, \\
 10144.4 & 10.16 &  9.11 &  8.73 &  8.2\,\,\, \\
 10320.7 & 10.27 &  9.25 &  8.82 &  8.36 \\
 10446.5 & 10.84 &  9.85 &  9.32 &       \\
 10475.4 & 10.87 &  9.88 &  9.36 &       \\
 10792.6 & 10.10 &  9.04 &  8.66 &  8.05 \\
\multicolumn{5}{l}{IRAS\,04407--7000}\\
  6704.6 & 10.69 &  9.25 &  8.50 &       \\
  6723.4 & 10.71 &  9.32 &  8.55 &  7.61 \\
  6750.5 & 10.79 &  9.39 &  8.64 &  7.70 \\
  6796.4 & 10.91 &  9.46 &  8.69 &       \\
  6866.3 & 11.05 &  9.60 &  8.85 &       \\
  7081.5 & 11.70 & 10.21 &  9.27 &       \\
  7098.5 & 11.77 & 10.29 &  9.33 &  8.23 \\
  7131.5 & 11.94 & 10.42 &  9.41 &  8.36 \\
  7153.4 & 11.99 & 10.52 &  9.49 &       \\
  7368.6 & 11.77 & 10.22 &  9.20 &  8.03 \\
  7495.5 & 11.11 &  9.59 &  8.69 &  7.57 \\
  9021.3 & 10.62 &  9.07 &  8.23 &  7.25 \\ 
  9297.5 & 11.25 &  9.74 &  8.86 &  7.90 \\
  9355.3 & 11.41 &  9.92 &  9.01 &  8.0\,  \\
  9434.2 & 11.80 & 10.22 &  9.23 &  8.0\,  \\
  9492.2 & 12.06 & 10.39 &  9.32 &  8.19 \\
  9582.7 & 12.39 & 10.58 &  9.43 &  8.2\,\,\, \\
  9744.4 & 11.68 & 10.07 &  8.98 &  7.8\,\,\, \\
  9791.3 & 11.12 &  9.64 &  8.70 &  7.47 \\
  9959.6 & 10.44 &  8.99 &  8.20 &  7.19 \\
 10023.6 & 10.37 &  8.92 &  8.15 &  7.08 \\
 10034.5 & 10.32 &  8.86 &  8.12 &  7.01 \\
\hline
\end{tabular}
\end{center}
\end{table}
\setcounter{table}{0}
\begin{table}
\begin{center}
\begin{tabular}{rrrrr}
\hline
  JD& $J$ & $H$ & $K$ & $L$ \\ 
--2440000& \multicolumn{4}{c}{(mag)}\\
\hline
\multicolumn{5}{l}{IRAS\,04407--7000 continued}\\
 10119.3 & 10.26 &  8.81 &  8.10 &  7.09 \\
 10317.6 & 10.62 &  9.16 &  8.43 &  7.58 \\
 10353.6 & 10.69 &  9.27 &  8.51 &  7.53 \\
 10390.5 & 10.79 &  9.40 &  8.61 &  7.70 \\
 10499.4 & 11.20 &  9.78 &  8.93 &  7.93 \\
\multicolumn{5}{l}{IRAS\,04498--6842}\\
  9294.5 &  9.69 &  8.61 &  7.92 &  6.94 \\
  9298.5 &  9.57 &  8.52 &  7.89 &  6.85 \\
  9300.5 &  9.55 &  8.53 &  7.87 &  6.88 \\
  9352.3 &  9.48 &  8.43 &  7.80 &  6.70 \\
  9434.3 &  9.34 &  8.20 &  7.64 &  6.71 \\
  9496.2 &  9.21 &  8.08 &  7.52 &  6.56 \\
  9586.6 &  9.13 &  7.96 &  7.43 &  6.57 \\
  9701.4 &  9.21 &  8.05 &  7.52 &  6.66 \\
  9744.4 &  9.29 &  8.14 &  7.64 &  6.94 \\
  9791.3 &  9.36 &  8.24 &  7.69 &  6.84 \\
  9960.6 &  9.99 &  8.83 &  8.20 &  7.46 \\
 10029.4 & 10.30 &  9.13 &  8.42 &  7.65 \\
 10143.3 & 10.73 &  9.51 &  8.67 &  7.70 \\
 10326.5 & 11.02 &  9.65 &  8.71 &  7.68 \\
 10390.5 & 11.13 &  9.78 &  8.77 &  7.65 \\
 10504.3 & 10.00 &  8.98 &  8.18 &  7.12 \\
 10712.6 &  9.38 &  8.28 &  7.67 &  6.76 \\
 10915.2 &  9.22 &  8.08 &  7.50 &  6.58 \\
 11210.4 &  9.96 &  8.82 &  8.18 &  7.4\,\,\, \\
 11626.2 & 10.62 &  9.33 &  8.52 &  7.50 \\
\multicolumn{5}{l}{IRAS\,04509--6922}\\
  9748.4 &  9.93 &  8.61 &  7.95 &  7.00 \\
  9791.3 &  9.95 &  8.64 &  7.97 &  7.03 \\
  9965.6 & 10.40 &  9.08 &  8.38 &  7.53 \\
 10031.4 & 10.65 &  9.34 &  8.59 &  7.65 \\
 10119.3 & 10.98 &  9.63 &  8.80 &  7.8\,\,\, \\
 10317.6 & 11.84 & 10.31 &  9.30 &  8.23 \\
 10448.5 & 11.84 & 10.32 &  9.29 &  8.0\,\,\, \\
 10500.4 & 11.70 & 10.25 &  9.22 &  8.0\,\,\, \\
 10764.4 & 10.18 &  8.87 &  8.14 &  7.16 \\
 10915.3 &  9.98 &  8.65 &  7.95 &  7.08 \\
\multicolumn{5}{l}{IRAS\,04516--6902}\\
  7495.6 & 12.27 & 10.42 &  9.28 &  8.01 \\
  9748.4 & 11.32 &  9.87 &  9.00 &  7.76 \\
  9791.4 & 10.82 &  9.47 &  8.66 &  7.48 \\
 10031.4 & 10.37 &  8.84 &  8.11 &  7.04 \\
 10147.2 & 10.39 &  8.84 &  8.12 &  7.14 \\
 10325.6 & 10.93 &  9.41 &  8.66 &  7.74 \\
 10392.5 & 11.19 &  9.73 &  8.92 &  8.10 \\
 10500.4 & 11.64 & 10.16 &  9.24 &  8.2\,\,\, \\
 10764.4 & 11.60 & 10.12 &  9.23 &  8.17 \\
 10915.3 & 10.51 &  9.19 &  8.43 &  7.26 \\
 11482.6 & 11.08 &  9.68 &  8.87 &  7.88 \\
\multicolumn{5}{l}{IRAS\,04545--7000}\\
  9797.3 &       & 14.05 & 11.20 &  8.6\,\,\, \\
 10031.6 &       & 12.09 &  9.71 &  7.34 \\
 10033.5 &       & 12.15 &  9.73 &  7.37 \\
 10123.4 &       & 11.77 &  9.41 &  7.11 \\
 10320.6 &       & 11.80 &  9.47 &  7.30 \\
 10448.5 &       & 12.47 &  9.79 &       \\
 10504.3 &       & 12.32 &  9.96 &  7.74 \\
 10761.5 &       & 13.06 & 10.60 &  8.3\,\,\, \\
 10796.4 &       & 13.14 & 10.70 &  8.4\,\,\, \\
 10887.3 &       & 13.54 & 10.96 &  8.6\,\,\, \\
 11095.5 &       & 12.82 & 10.38 &  8.0\,\,\, \\
 11624.2 &       & 11.70 &  9.56 &  7.44 \\
 11834.6 &       & 12.29 & 10.08 &  8.08 \\
\hline
\end{tabular}
\end{center}
\end{table}
\setcounter{table}{0}
\begin{table}
\begin{center}
\caption[]{continued}
\begin{tabular}{rrrrr}
\hline
  JD& $J$ & $H$ & $K$ & $L$ \\ 
--2440000& \multicolumn{4}{c}{(mag)}\\
\hline
\multicolumn{5}{l}{IRAS\,05003--6712}\\
  9022.3 & 12.62 & 11.05 &  9.94 &       \\ 
  9742.3 & 14.2\,\,\, & 12.41 & 10.85 &       \\
  9959.7 & 12.36 & 10.82 &  9.54 &  8.00 \\
 10027.6 & 12.11 & 10.59 &  9.32 &  7.82 \\
 10123.3 & 12.06 & 10.45 &  9.24 &  7.86 \\
 10325.5 & 12.77 & 11.19 &  9.94 &  8.46 \\
 10392.5 & 12.96 & 11.46 & 10.15 &  8.85 \\
 10474.5 & 13.4\,\,\, & 11.79 & 10.43 &  9.0\,\,\, \\
 10764.4 & 13.26 & 11.53 & 10.13 &  8.6\,\,\, \\
 10796.4 & 12.82 & 11.13 &  9.83 &  8.3\,\,\, \\
 10915.3 & 12.23 & 10.52 &  9.31 &  7.86 \\
 11095.5 & 12.14 & 10.40 &  9.29 &  8.0\,\,\, \\
\multicolumn{5}{l}{IRAS\,05294--7104}\\
  9964.6 & 11.38 &  9.68 &  8.70 &  7.4\,\,\, \\
 10032.5 & 11.46 &  9.74 &  8.74 &  7.57 \\
 10327.6 & 12.57 & 10.70 &  9.57 &  8.44 \\
 10501.4 & 13.36 & 11.27 &  9.94 &  8.4\,\,\, \\
 10705.6 & 12.20 & 10.32 &  9.19 &  7.92 \\
 10889.3 & 11.47 &  9.70 &  8.66 &  7.38 \\
\multicolumn{5}{l}{IRAS\,05329--6708}\\
  9024.5 &       &       & 11.65 &       \\ 
  9298.4 &       & 11.99 &  9.65 &  7.38 \\
  9747.5 &       & 11.61 &  9.42 &  7.26 \\
 10033.5 &       & 12.55 & 10.27 &  8.13 \\
 10446.5 &       & 13.6\,\,\, & 10.36 &  7.9\,\,\, \\
 10501.5 &       & 12.17 &  9.83 &  7.48 \\
 10797.5 &       & 11.20 &  9.07 &  6.84 \\
 10887.4 & 15.2\,\,\, & 11.25 &  9.14 &  6.95 \\
\multicolumn{5}{l}{IRAS\,05402--6956}\\
 10035.6 &       &       & 11.61 &  8.8\,\,\, \\
 10327.6 &       & 12.52 & 10.10 &  7.68 \\
 10447.5 &       & 11.88 &  9.67 &  7.45 \\
 10504.4 &       & 11.71 &  9.59 &  7.39 \\
 10706.6 & 14.3\,\,\, & 11.33 &  9.35 &  7.15 \\
 10915.3 &       & 11.86 &  9.79 &  7.59 \\
 11210.5 &       &       & 11.6\,\,\, &       \\
\multicolumn{5}{l}{IRAS\,05558--7000}\\
  9024.5 & 11.80 &  9.93 &  8.84 &  7.46 \\ 
  9298.5 & 12.83 & 10.88 &  9.63 &  8.43 \\
  9355.4 & 13.03 & 11.07 &  9.75 &  8.3\,\,\, \\
  9439.3 & 13.4\,\,\, & 11.29 &  9.88 &  8.5\,\,\, \\
  9492.2 & 13.45 & 11.30 &  9.88 &  8.5\,\,\, \\
  9702.5 & 12.38 & 10.47 &  9.15 &  7.71 \\
  9742.5 & 12.14 & 10.18 &  8.96 &  7.6\,\,\,\\
  9797.2 & 11.89 &  9.96 &  8.80 &  7.3\,\,\, \\
  9961.6 & 11.70 &  9.73 &  8.59 &  7.32 \\
 10027.6 & 11.65 &  9.71 &  8.58 &  7.26 \\
 10142.5 & 11.70 &  9.84 &  8.71 &  7.36 \\
 10325.7 & 12.53 & 10.68 &  9.44 &  8.06 \\
 10390.6 & 12.87 & 10.99 &  9.68 &  8.20 \\
 10476.5 & 13.4\,\,\, & 11.35 &  9.94 &       \\
 10798.5 & 13.4\,\,\, & 11.47 & 10.01 &  8.1\,\,\, \\
 10915.3 & 12.56 & 10.75 &  9.38 &  7.70 \\
\multicolumn{5}{l}{SHV\,04544--6848   (SP\,30-6)}\\
  9747.3 & 11.27 & 10.27 &  9.55 &  8.5\,\,\, \\
  9960.6 & 10.24 &  9.22 &  8.76 &  8.05 \\
 10033.5 & 10.19 &  9.19 &  8.67 &  8.06 \\
 10120.3 & 10.28 &  9.24 &  8.76 &  8.2\,\,\, \\
 10320.6 & 11.39 & 10.29 &  9.64 &  8.9\,\,\, \\
 10446.4 & 10.94 &  9.96 &  9.36 &       \\
 10472.5 & 10.96 &  9.97 &  9.39 &       \\
\hline
\end{tabular}
\end{center}
\end{table}
\setcounter{table}{0}
\begin{table}
\begin{center}
\begin{tabular}{rrrrr}
\hline
  JD& $J$ & $H$ & $K$ & $L$ \\ 
--2440000& \multicolumn{4}{c}{(mag)}\\
\hline
\multicolumn{5}{l}{SHV\,05220--7012}\\
  9797.4 & 13.40 & 12.48 & 12.05 &       \\
 10029.5 & 13.43 & 12.40 & 12.02 &       \\
 10147.3 & 12.94 & 11.97 & 11.53 &       \\
 10324.7 & 12.66 & 11.65 & 11.39 &       \\
 10442.5 & 13.38 & 12.47 & 12.08 &       \\
 10474.5 & 13.2\,\,\, & 12.40 & 12.01 &       \\
 10498.4 & 12.89 & 11.92 & 11.64 &       \\
\multicolumn{5}{l}{SHV\,05249--6945}\\
  9963.6 & 11.61 & 10.53 & 10.23 &       \\
 10029.5 & 11.71 & 10.57 & 10.16 &       \\
 10144.4 & 12.58 & 11.33 & 10.68 &       \\
 10323.7 & 11.92 & 10.91 & 10.58 &       \\
 10442.5 & 11.69 & 10.52 & 10.10 &       \\
 10504.4 & 12.18 & 10.91 & 10.32 &       \\
\multicolumn{5}{l}{SHV\,05305--7022}\\
  9965.5 & 12.22 & 11.32 & 10.91 &       \\
 10034.4 & 11.91 & 11.02 & 10.64 &       \\
 10142.4 & 11.68 & 10.66 & 10.25 &       \\
 10323.6 & 12.23 & 11.34 & 10.86 &       \\
 10383.6 & 12.03 & 11.15 & 10.81 &       \\
 10442.5 & 11.70 & 10.66 & 10.33 &       \\
 10475.4 & 11.67 & 10.61 & 10.23 &       \\
 10504.4 & 11.71 & 10.67 & 10.27 &       \\
\multicolumn{5}{l}{R105}\\
 10387.6 & 11.92 & 10.90 & 10.52 &       \\
 10498.3 & 11.45 & 10.41 & 10.07 &       \\
 10602.2 & 11.74 & 10.62 & 10.17 &       \\
 10762.5 & 12.00 & 10.95 & 10.57 &       \\
\multicolumn{5}{l}{WBP\,74}\\
 10443.4 & 12.96 & 11.89 & 11.46 &       \\
 10470.5 & 13.13 & 12.11 & 11.65 &       \\
 10498.3 & 13.13 & 12.09 & 11.76 &       \\
 10602.3 & 12.72 & 11.70 & 11.26 &       \\
 10705.6 & 13.18 & 12.13 & 11.71 &       \\
 10761.6 & 12.85 & 12.02 & 11.53 &       \\
\multicolumn{5}{l}{GRV\,0517584--655140}\\
 10443.4 & 13.39 & 12.46 & 12.2\,\,\, &       \\
 10470.4 & 13.02 & 12.20 & 11.91 &       \\
 10498.3 & 13.41 & 12.56 & 12.20 &       \\
 10602.2 & 13.25 & 12.46 & 12.14 &       \\
 10760.6 & 13.70 & 12.85 & 12.5\,\,\, &       \\
\multicolumn{5}{l}{RHV\,0524173--660913}\\
  8227.5 & 10.80 &  9.79 &  9.37 &       \\
 10147.2 & 10.62 &  9.63 &  9.24 &       \\
 10202.3 & 10.87 &  9.89 &  9.47 &       \\
 10324.7 & 11.43 & 10.48 & 10.00 &       \\
 10442.4 & 11.36 & 10.44 & 10.05 &       \\
 10498.3 & 10.99 & 10.05 &  9.71 &       \\
 10602.3 & 10.62 &  9.62 &  9.28 &       \\
 10855.4 & 11.32 & 10.46 & 10.01 &       \\
 10884.3 & 11.35 & 10.46 & 10.02 &       \\
 10915.3 & 11.41 & 10.50 & 10.08 &  9.24 \\
\multicolumn{5}{l}{\it C-rich AGB stars}\\
\multicolumn{5}{l}{IRAS\,04286--6937}\\
  9020.3 &       & 13.58 & 11.61 &  9.5\,\,\, \\
  9294.4 &       & 12.32 & 10.48 &       \\
  9298.4 &       & 12.28 & 10.53 &  8.4\,\,\, \\
  9742.4 &       & 12.93 & 11.17 &       \\
  9938.6 & 14.2\,\,\, & 12.02 & 10.31 &  8.32 \\
 10024.4 & 14.2\,\,\, & 12.11 & 10.42 &  8.3\,\,\, \\
 10034.5 & 14.5\,\,\, & 12.13 & 10.40 &  8.4\,\,\, \\
 10144.3 &       & 12.81 & 11.01 &  8.8\,\,\, \\
 10324.6 &       & 13.20 & 11.40 &  9.3\,\,\, \\
\hline
\end{tabular}
\end{center}
\end{table}
\setcounter{table}{0}
\begin{table}
\begin{center}
\caption[]{continued}
\begin{tabular}{rrrrr}
\hline
  JD& $J$ & $H$ & $K$ & $L$ \\ 
--2440000& \multicolumn{4}{c}{(mag)}\\
\hline
\multicolumn{5}{l}{IRAS\,04286--6937 continued}\\
 10387.5 &       & 13.20 & 11.34 &  9.0\,\,\, \\
 10500.3 &       & 12.44 & 10.63 &  8.6\,\,\, \\
 11626.2 &       & 13.6\,\,\, & 11.58 &       \\
 11890.4 &       & 12.42 & 10.58 &  8.4\,\,\, \\
\multicolumn{5}{l}{IRAS\,04374--6831}\\
  9020.3 &       & 14.3\,\,\, & 12.01 &  9.3\,\,\, \\
  9295.3 &       &       & 12.24 &       \\
  9745.3 &       &       & 12.4\,\,\, &  9.6\,\,\, \\
 10027.4 &       & 13.6\,\,\, & 11.47 &  9.1\,\,\, \\
 10034.5 &       & 13.7\,\,\, & 11.41 &  9.0\,\,\, \\
 10144.3 &       & 13.5\,\,\, & 11.29 &  8.78 \\
 10326.5 &       &       & 12.20 &       \\
 10388.6 &       &       & 12.57 &       \\
 10500.4 &       &       & 12.7\,\,\, &       \\
 10765.4 &       & 13.7\,\,\, & 11.34 &       \\
 10915.2 &       &       & 11.85 &  9.0\,\,\, \\
 11626.3 &       &       & 12.34 &       \\
\multicolumn{5}{l}{IRAS\,04496--6958}  \\                                                
  9021.3 & 12.10 & 10.19 &  8.80 &  7.25 \\
  9115.2 & 12.42 & 10.43 &  8.98 &  7.45 \\
  9295.4 & 13.0\,\,\, & 11.02 &  9.53 &  7.9\,\,\, \\
  9353.3 & 13.06 & 11.12 &  9.61 &  8.0\,\,\, \\
  9434.3 & 13.14 & 11.15 &  9.68 &  8.00 \\
  9495.2 & 12.82 & 10.84 &  9.39 &  7.82 \\
  9586.6 & 12.36 & 10.42 &  9.02 &  7.49 \\
  9701.4 & 12.08 & 10.20 &  8.80 &  7.3\,\,\, \\
  9744.4 & 12.09 & 10.22 &  8.82 &  7.29 \\
  9791.3 & 12.23 & 10.31 &  8.91 &  7.35 \\
  9959.6 & 12.77 & 10.81 &  9.33 &  7.72 \\
 10029.4 & 13.05 & 11.03 &  9.53 &  7.89 \\
 10124.3 & 13.13 & 11.16 &  9.62 &  7.81 \\
 10324.6 & 12.22 & 10.34 &  8.91 &  7.50 \\
 10391.5 & 11.94 & 10.05 &  8.68 &  7.25 \\
 10446.4 & 11.95 & 10.04 &  8.68 &  7.30 \\
 10504.3 & 12.09 & 10.11 &  8.74 &  7.36 \\
 10856.3 & 12.93 & 10.96 &  9.47 &  7.8\,\,\, \\
 11625.3 & 12.98 & 11.03 &  9.56 &  7.99 \\
 11888.5 & 11.89 & 10.05 &  8.73 &  7.23 \\
 11982.2 & 12.20 & 10.34 &  8.97 &  7.43 \\
\multicolumn{5}{l}{IRAS\,04539--6821}\\
  9021.4 &       &       & 12.19 &  9.14 \\
  9746.4 &       &       & 12.67 &  9.67 \\
  9961.6 &       &       & 13.1\,\,\,  &       \\
 10031.4 &       &       & 12.9\,\,\,  &       \\
 10147.3 &       &       & 11.88 &       \\
 10327.5 &       &       & 11.98 &  9.05 \\
 10387.5 &       &       & 12.35 &  9.56 \\
 10500.4 &       &       & 13.1\,\,\,  &       \\
 10764.5 &       &       & 12.4\,\,\,  &  9.44 \\
\multicolumn{5}{l}{IRAS\,04557--6753}\\
  9021.3 &       & 13.24 & 11.22 &  8.75 \\
  9116.2 &       & 13.47 & 11.23 &  8.69 \\
  9294.5 &       &       & 11.9\,\,\, &  9.0\,\,\, \\
  9298.6 &       &       & 11.96 &       \\
  9352.4 &       & 14.4\,\,\, & 12.23 &  9.4\,\,\, \\
  9434.3 &       &       & 12.4\,\,\, &       \\
  9701.5 &       & 13.6\,\,\, & 11.47 &       \\
  9745.3 &       & 13.6\,\,\, & 11.30 &  8.8\,\,\, \\
  9746.4 &       & 13.33 & 11.28 &  8.8\,\,\, \\
  9791.4 &       & 13.4\,\,\, & 11.25 &  8.7\,\,\, \\
  9959.6 &       & 13.43 & 11.36 &  8.8\,\,\, \\
 10028.4 &       & 13.7\,\,\, & 11.66 &  9.0\,\,\, \\
\hline
\end{tabular}
\end{center}
\end{table}
\setcounter{table}{0}
\begin{table}
\begin{center}
\begin{tabular}{rrrrr}
\hline
  JD& $J$ & $H$ & $K$ & $L$ \\ 
--2440000& \multicolumn{4}{c}{(mag)}\\
\hline
\multicolumn{5}{l}{IRAS\,04557--6753 continued}\\
 10144.3 &       & 14.5\,\,\, & 12.32 &       \\
 10327.5 &       &       & 12.39 &  9.6\,\,\, \\
 10500.4 &       & 13.4\,\,\, & 11.19 &  8.9\,\,\, \\
 11626.3 &       &       & 12.10 &       \\
\multicolumn{5}{l}{IRAS\,05009--6616}\\
  9027.4 & 14.6\,\,\,  & 12.50 & 10.76 &  8.9\,\,\,  \\
  9296.4 &       & 13.3\,\,\,  & 11.59 &       \\
  9352.4 &       & 13.3\,\,\,  & 11.46 &  9.1\,\,\,  \\
  9434.3 &       & 12.63 & 10.90 &  8.6\,\,\,  \\
  9495.2 &       & 12.39 & 10.61 &  8.37 \\
  9586.7 &       & 12.26 & 10.57 &  8.37 \\
  9701.5 &       & 12.90 & 11.12 &  8.8\,\,\,  \\
  9742.4 &       & 13.4\,\,\,  & 11.55 &       \\
  9791.4 &       & 13.6\,\,\,  & 11.79 &       \\
  9960.7 &       & 13.7\,\,\,  & 11.92 &  9.5\,\,\,  \\
 10027.4 &       & 13.6\,\,\,  & 11.75 &  9.2\,\,\,  \\
 10123.4 &       & 12.78 & 10.94 &  8.6\,\,\,  \\
 10325.5 &       & 12.83 & 11.06 &  8.6\,\,\,  \\
 10390.4 &       & 13.17 & 11.51 &  9.1\,\,\,  \\
 10475.3 &       & 13.9\,\,\,  & 12.0\,\,\,  &       \\
 10500.5 &       & 13.7\,\,\,  & 12.04 &       \\
 10764.5 &       & 12.96 & 11.04 &  8.6\,\,\,  \\
 10792.5 &       & 12.82 & 11.01 &  8.6\,\,\,  \\
 11210.4 &       & 14.4\,\,\,  & 12.57 &       \\
 11480.5 &       & 13.2\,\,\,  & 11.14 &  8.56\,\,\,  \\
 11625.3 &       & 13.5\,\,\,  & 11.48 &  8.6\,\,\,  \\
 11888.5 &       &       & 12.75 &       \\
 11988.4 &       &       & 12.53 &  9.5\,\,\, \\
\multicolumn{5}{l}{IRAS\,05112--6755}\\
  9021.5 &       &       & 12.7\,\,\,  &  9.4\,\,\,  \\
  9297.3 &       &       & 13.2\,\,\,  &       \\
  9747.4 &       & 14.3\,\,\,  & 12.07 &  8.9\,\,\,  \\
 10032.5 &       &       & 13.1\,\,\,  &       \\
 10328.5 &       & 14.2\,\,\,  & 11.69 &  8.6\,\,\,  \\
 10387.5 &       & 14.0\,\,\,  & 11.55 &  8.5\,\,\,  \\
 10498.4 &       & 14.1\,\,\,  & 11.59 &  8.5\,\,\,  \\
 10764.5 &       &       & 13.0\,\,\,  &  9.6\,\,\,  \\
\multicolumn{5}{l}{IRAS\,05113--6739}\\
  9022.4 &       &       & 13.1\,\,\,  &       \\
  9297.4 &       &       & 12.9\,\,\,  &       \\
  9745.4 &       &       & 12.97 &  9.8\,\,\,  \\
 10032.4 &       &       & 12.5\,\,\,  &       \\
 10328.5 &       & 14.2\,\,\,  & 12.05 &  9.0\,\,\,  \\
 10796.5 &       & 14.0\,\,\,  & 11.68 &  8.9\,\,\,  \\
 10915.3 &       & 13.8\,\,\,  & 11.43 &  8.56 \\
\multicolumn{5}{l}{IRAS\,05128--6455}\\
  9022.4 & 13.31 & 11.69 & 10.31 &  8.5\,\,\, \\
  9116.2 & 13.44 & 11.92 & 10.52 &  8.74 \\
  9296.4 &       & 13.2\,\,\, & 11.41 &       \\
  9355.4 &       & 13.1\,\,\, & 11.44 &       \\
  9434.3 &       & 13.2\,\,\, & 11.53 &       \\
  9496.2 &       & 12.9\,\,\, & 11.29 &  9.2\,\,\, \\
  9701.6 &       & 11.86 & 10.36 &  8.4\,\,\, \\
  9742.4 & 13.51 & 11.80 & 10.38 &  8.42 \\
  9792.3 & 13.60 & 11.89 & 10.47 &  8.6\,\,\, \\
  9938.7 &       & 12.32 & 10.87 &  9.1\,\,\, \\
 10023.6 &       & 12.74 & 11.22 &       \\
 10034.4 & 14.6\,\,\, & 12.81 & 11.24 &  9.4\,\,\, \\
 10126.5 &       & 13.14 & 11.48 &       \\
 10325.6 &       & 12.37 & 10.75 &  8.65 \\
 10473.5 &       & 11.98 & 10.50 &  8.4\,\,\, \\
 10917.2 &       & 12.72 & 10.99 &       \\
 11625.3 &       & 12.90 & 11.29 &       \\
\hline
\end{tabular}
\end{center}
\end{table}
\setcounter{table}{0}
\begin{table}
\begin{center}
\caption[]{continued}
\begin{tabular}{rrrrr}
\hline
  JD& $J$ & $H$ & $K$ & $L$ \\ 
--2440000& \multicolumn{4}{c}{(mag)}\\
\hline
\multicolumn{5}{l}{IRAS\,05190--6748}\\
  9057.3 &       &       & 12.18 &  8.6\,\,\, \\
  9117.2 &       &       & 12.37 &  8.86 \\
  9297.4 &       &       & 13.3\,\,\, &  9.8\,\,\, \\
  9742.5 &       &       & 12.33 &  8.9\,\,\, \\
 10031.5 &       &       & 12.00 &  8.8\,\,\, \\
 10125.3 &       &       & 12.4\,\,\, &       \\
 10326.5 &       &       & 13.3\,\,\, &       \\
 10390.5 &       &       & 13.4\,\,\, &       \\
 10501.4 &       &       & 13.5\,\,\, &       \\
 10764.6 &       &       & 12.16 &  8.7\,\,\, \\
\multicolumn{5}{l}{IRAS\,05291--6700}\\
  9021.5 & 12.8\,\,\, & 11.02 &  9.93 &  8.8\,\,\, \\
  9296.5 & 13.6\,\,\, & 11.75 & 10.45 &  8.9\,\,\, \\
  9355.4 & 13.02 & 11.47 & 10.30 &  8.9\,\,\, \\
  9439.3 & 12.83 & 11.26 & 10.16 &  8.9\,\,\, \\
  9495.2 & 12.59 & 11.05 & 10.01 &  8.62 \\
  9587.6 & 12.61 & 10.92 &  9.86 &  8.7\,\,\, \\
  9702.4 & 13.49 & 11.69 & 10.38 &  9.0\,\,\, \\
  9747.4 & 13.44 & 11.61 & 10.34 &  9.0\,\,\, \\
\multicolumn{5}{l}{IRAS\,05295--7121}\\
  9023.3 &       & 12.88 & 11.03 &  9.0\,\,\, \\
  9115.2 & 14.3\,\,\, & 12.17 & 10.42 &  8.5\,\,\, \\
  9296.6 &       & 12.32 & 10.49 &  8.4\,\,\, \\
  9355.4 &       & 12.46 & 10.66 &  8.5\,\,\, \\
  9439.3 &       & 13.1\,\,\, & 11.18 &  9.1\,\,\, \\
  9496.2 &       & 13.5\,\,\, & 11.51 &  9.2\,\,\, \\
  9702.5 &       & 13.2\,\,\, & 11.14 &  8.9\,\,\, \\
  9744.5 &       & 12.8\,\,\, & 10.82 &  8.6\,\,\, \\
  9792.4 &       & 12.49 & 10.63 &  8.5\,\,\, \\
  9961.6 &       & 12.68 & 10.67 &  8.6\,\,\, \\
 10027.5 &       & 13.27 & 11.08 &  8.9\,\,\, \\
 10325.7 &       & 13.6\,\,\, & 11.53 &  9.2\,\,\, \\
 10390.5 &       & 13.07 & 11.00 &  8.7\,\,\, \\
 10504.4 &       & 12.71 & 10.69 &  8.6\,\,\, \\
\multicolumn{5}{l}{IRAS\,05300--6651}\\
  9023.4 &       & 13.8\,\,\, & 11.58 &  9.0\,\,\, \\
  9113.2 &       & 14.4\,\,\, & 12.02 &  9.3\,\,\, \\
  9297.5 &       &       & 13.0\,\,\, &       \\
  9745.5 &       & 14.3\,\,\, & 11.72 &  8.6\,\,\, \\
 10326.6 &       & 14.3\,\,\, & 11.94 &  8.9\,\,\, \\
 10501.5 &       &       & 12.17 &       \\
 10797.5 &       &       & 13.4\,\,\, &       \\
 10915.3 &       &       & 12.33 &       \\
\multicolumn{5}{l}{IRAS\,05360--6648}\\
  9026.5 &       &       & 13.0\,\,\, &       \\
  9300.6 &       & 13.7\,\,\, & 12.41 &  9.4\,\,\, \\
  9742.6 &       &       & 12.3\,\,\, &       \\
 10326.6 &       &       & 12.15 &       \\
 10327.6 &       &       & 12.27 &  9.2\,\,\, \\
 10446.5 &       &       & 12.7\,\,\, &       \\
 10504.4 &       &       & 13.2\,\,\, &       \\
\multicolumn{5}{l}{SHV\,05003-6817}\\
  9748.3 & 14.2\,\,\, & 12.17 & 10.84 &       \\
  9960.6 & 12.92 & 11.27 & 10.15 &  9.0\,\,\, \\
 10028.6 & 13.20 & 11.48 & 10.32 &       \\
 10144.3 & 14.4\,\,\, & 12.35 & 10.95 &       \\
 10324.6 & 13.06 & 11.43 & 10.27 &  9.0\,\,\, \\
 10355.6 & 13.06 & 11.38 & 10.22 &       \\
 10383.5 & 13.17 & 11.46 & 10.26 &  8.9\,\,\, \\
 10472.5 & 13.9\,\,\, & 12.13 & 10.80 &       \\
\hline
\end{tabular}
\end{center}
\end{table}
\setcounter{table}{0}
\begin{table}
\begin{center}
\begin{tabular}{rrrrr}
\hline
  JD& $J$ & $H$ & $K$ & $L$ \\ 
--2440000& \multicolumn{4}{c}{(mag)}\\
\hline
\multicolumn{5}{l}{SHV\,05003--6829  }\\                                               
  9748.3 & 13.06 & 11.25 &  9.91 &  8.4\,\,\, \\
  9960.6 & 13.87 & 11.97 & 10.48 &  9.0\,\,\, \\
 10031.5 & 14.0\,\,\, & 11.98 & 10.56 &  9.0\,\,\, \\
 10123.4 & 13.60 & 11.77 & 10.40 &  8.8\,\,\, \\
 10323.6 & 13.15 & 11.35 & 10.07 &  8.6\,\,\, \\
 10355.6 & 13.4\,\,\, & 11.57 & 10.24 &  8.9\,\,\, \\
 10382.6 & 13.6\,\,\, & 11.78 & 10.40 &       \\
 10443.5 & 14.1\,\,\, & 12.20 & 10.73 &       \\
 10474.4 & 14.2\,\,\, & 12.23 & 10.74 &  9.2\,\,\, \\
\multicolumn{5}{l}{SHV\,05027--6924}\\
  9748.3 & 12.96 & 11.66 & 10.98 &       \\
  9960.7 & 12.49 & 11.15 & 10.60 &       \\
 10028.5 & 13.03 & 11.70 & 10.89 &       \\
 10144.4 & 12.41 & 11.38 & 10.87 &       \\
 10324.6 & 12.97 & 11.65 & 10.89 &       \\
 10355.6 & 12.96 & 11.69 & 10.97 &       \\
 10498.3 & 12.19 & 11.09 & 10.71 &       \\
 10915.3 & 12.98 & 11.59 & 10.87 &       \\
\multicolumn{5}{l}{SHV\,05210--6904}\\
  9797.3 & 11.46 & 10.08 &  9.37 &  8.6\,\,\, \\
  9963.6 & 11.72 & 10.50 &  9.74 &       \\
 10031.5 & 11.55 & 10.33 &  9.68 &  8.9\,\,\, \\
 10126.4 & 11.22 & 10.06 &  9.50 &  8.7\,\,\, \\
 10320.7 & 11.35 &  9.98 &  9.32 &  8.7\,\,\, \\
 10447.6 & 12.08 & 10.75 &  9.84 &       \\
 10501.4 & 12.03 & 10.74 &  9.86 &       \\
 10798.4 & 11.07 &  9.77 &  9.23 &  8.4\,\,\, \\
 10887.3 & 11.49 & 10.08 &  9.39 &  8.8\,\,\, \\
 10917.3 & 11.75 & 10.31 &  9.53 &       \\
 11480.5 & 11.83 & 10.44 &  9.61 &  9.2\,\,\, \\
 11623.2 & 11.82 & 10.64 &  9.75 &       \\
 11888.5 & 10.92 &  9.67 &  9.06 &  8.3\,\,\, \\
\multicolumn{5}{l}{SHV\,05260--7011}\\                                              
  9963.6 &       & 12.37 & 10.90 &       \\
 10031.6 & 13.9\,\,\, & 12.06 & 10.72 &  9.4\,\,\, \\
 10142.4 & 13.20 & 11.37 & 10.18 &       \\
 10323.7 &       & 12.37 & 10.79 &       \\
 10442.5 & 13.7\,\,\, & 11.78 & 10.43 &       \\
 10476.4 & 13.9\,\,\, & 11.73 & 10.42 &       \\
 10797.4 & 13.9\,\,\, & 12.00 & 10.58 &  9.2\,\,\, \\
 10887.3 & 13.22 & 11.52 & 10.29 &  9.1\,\,\, \\
 10917.3 & 13.31 & 11.50 & 10.24 &  8.9\,\,\, \\
\multicolumn{5}{l}{WBP\,14}\\                                                      
  9963.6 & 13.5\,\,\, & 11.81 & 10.71 &       \\
 10028.5 & 13.8\,\,\, & 12.06 & 10.88 &       \\
 10142.4 & 13.25 & 11.66 & 10.61 &       \\
 10324.7 & 13.5\,\,\, & 11.76 & 10.64 &       \\
 10442.5 & 13.47 & 11.83 & 10.69 &       \\
 10475.4 & 13.33 & 11.69 & 10.61 &       \\
 10504.4 & 13.05 & 11.49 & 10.45 &       \\
 10797.4 & 13.5\,\,\, & 11.85 & 10.72 &       \\
 10887.3 & 13.02 & 11.42 & 10.47 &  9.5\,\,\, \\
 10915.3 & 12.96 & 11.39 & 10.40 &  9.3\,\,\, \\
\multicolumn{5}{l}{TRM\,72 (05116--6654)}\\
  9027.4 &       &       & 10.50 &       \\
  9058.2 &       & 12.55 & 10.67 &  8.34 \\
  9117.2 &       & 13.06 & 11.06 &  8.65 \\
  9295.5 &       &       & 11.96 &       \\
  9352.4 &       & 13.41 & 11.52 &  9.1\,\,\, \\
  9434.3 &       & 13.37 & 11.50 &  9.2\,\,\, \\
  9587.6 &       & 12.61 & 10.86 &  8.5\,\,\, \\
  9701.5 &       & 13.2\,\,\, & 11.58 &       \\
  9745.4 &       & 14.1\,\,\, & 12.02 &  9.4\,\,\, \\
  9792.3 &       & 14.4\,\,\, & 12.38 &       \\
\hline
\end{tabular}
\end{center}
\end{table}
\setcounter{table}{0}
\begin{table}
\begin{center}
\caption[]{continued}
\begin{tabular}{rrrrr}
\hline
  JD& $J$ & $H$ & $K$ & $L$ \\ 
--2440000& \multicolumn{4}{c}{(mag)}\\
\hline
\multicolumn{5}{l}{TRM\,72 continued}\\  
  9938.7 &       & 14.0\,\,\, & 11.95 &  9.3\,\,\, \\
 10027.5 &       & 13.32 & 11.45 &  9.0\,\,\, \\
 10122.4 &       & 13.07 & 11.07 &  8.5\,\,\, \\
 10201.2 &       & 13.5\,\,\, & 11.41 &  8.7\,\,\, \\
 10325.4 &       &       & 12.28 &  9.4\,\,\, \\
 10388.6 &       &       & 12.44 &  9.6\,\,\, \\
 10501.4 &       & 14.2\,\,\, & 11.90 &       \\
 10764.6 &       & 12.65 & 10.89 &  8.6\,\,\, \\
 10792.5 &       & 12.84 & 10.98 &  8.6\,\,\, \\
 10915.3 &       & 13.7\,\,\, & 11.52 &  9.3\,\,\, \\
 11210.4 & 13.8\,\,\, & 11.69 & 10.12 &  8.2\,\,\, \\
 11480.5 &       & 12.99 & 11.21 &  9.1\,\,\, \\
 11625.3 &       & 12.65 & 11.02 &  8.8\,\,\, \\
 11888.5 &       & 12.40 & 10.67 &  8.6\,\,\, \\
 11988.4 &       & 13.25 & 11.39 &  9.0\,\,\, \\
\multicolumn{5}{l}{TRM\,88   (05202--6638)}\\
  9057.3 &       & 13.60 & 11.39 &  9.0\,\,\, \\
  9117.2 &       & 13.07 & 11.13 &  9.0\,\,\, \\
  9296.5 &       & 12.20 & 10.47 &       \\
  9300.5 &       & 12.13 & 10.43 &  8.59 \\
  9352.4 &       & 12.16 & 10.40 &  8.5\,\,\, \\
  9439.3 &       & 12.49 & 10.65 &  8.68 \\
  9702.4 &       & 12.01 & 10.40 &  8.7\,\,\, \\
  9742.5 & 13.9\,\,\, & 11.75 & 10.18 &  8.6\,\,\, \\
  9792.3 & 13.11 & 11.24 &  9.83 &  8.3\,\,\, \\
  9961.6 & 13.35 & 11.26 &  9.86 &  8.30 \\
 10027.5 & 13.7\,\,\, & 11.59 & 10.14 &  8.5\,\,\, \\
 10142.3 & 13.7\,\,\, & 11.66 & 10.19 &  8.4\,\,\, \\
 10201.2 & 13.4\,\,\, & 11.45 & 10.06 &  8.6\,\,\, \\
 10325.6 & 12.6\,\,\, & 10.87 &  9.62 &  8.23 \\
 10390.6 & 12.44 & 10.71 &  9.46 &  8.05 \\
 10499.3 & 13.08 & 11.12 &  9.73 &  8.2\,\,\, \\
 10764.6 &       & 11.77 & 10.25 &  8.6\,\,\, \\
 10792.6 & 13.6\,\,\, & 11.78 & 10.24 &  8.5\,\,\, \\
 11210.4 &       & 13.0\,\,\, & 11.23 &       \\
 11625.3 &       & 13.2\,\,\, & 11.34 &  9.0\,\,\, \\
 11888.5 &       & 13.51 & 11.55 &  9.3\,\,\, \\
 11988.4 &       & 12.62 & 10.82 &  8.7\,\,\, \\
\multicolumn{5}{l}{\it Supergiants, SRs and others}\\
\multicolumn{5}{l}{HV\,5870}\\
  4282.4 &  9.0\,\,\, &  8.1\,\,\, &  7.8\,\,\, &   *   \\
  4552.5 &  9.3\,\,\, &  8.43 &  8.06 &    *  \\
  4923.6 &  9.31 &  8.36 &  7.98 &   *   \\
  5012.4 &  9.34 &  8.41 &  8.05 &   *   \\
  5259.5 &  9.18 &  8.31 &  7.99 &   *   \\
  5649.5 &  9.34 &  8.47 &  8.17 &   *   \\
  5688.4 &  9.28 &  8.40 &  8.11 &   *   \\
  6031.4 &  9.53 &  8.63 &  8.24 &   *   \\
  6357.5 &  9.27 &  8.36 &  8.06 &   *   \\
  6381.6 &  9.23 &  8.34 &  8.04 &   *   \\
  6425.5 &  9.17 &  8.30 &  8.00 &   *   \\
  6489.4 &  9.19 &  8.31 &  7.97 &   *   \\
  6784.4 &  9.78 &  8.80 &  8.44 &   *   \\
  9705.4 &  9.26 &  8.32 &  7.97 &  7.48 \\
  9965.5 &  9.41 &  8.47 &  8.13 &  7.62 \\
 10033.5 &  9.39 &  8.47 &  8.14 &  7.62 \\
 10121.4 &  9.35 &  8.41 &  8.12 &  7.6\,\,\, \\
 10323.7 &  9.12 &  8.18 &  7.85 &  7.34 \\
 10447.5 &  9.24 &  8.27 &  7.93 &  7.5\,\,\, \\
 10501.5 &  9.35 &  8.39 &  8.01 &  7.47 \\
 10798.4 &  9.29 &  8.35 &  8.03 &  7.53 \\
 10858.5 &  9.34 &  8.39 &  8.03 &  7.6\,\,\, \\
\hline
\end{tabular}
\end{center}
\end{table}
\setcounter{table}{0}
\begin{table}
\begin{center}
\begin{tabular}{rrrrr}
\hline
  JD& $J$ & $H$ & $K$ & $L$ \\ 
--2440000& \multicolumn{4}{c}{(mag)}\\
\hline
\multicolumn{5}{l}{IRAS\,04530--6916}\\                                                  
  9797.3 & 13.7\,\,\, & 11.73 &  9.86 &  7.5\,\,\, \\
 10032.5 & 13.6\,\,\, & 11.58 &  9.71 &  7.5\,\,\, \\
 10120.3 & 13.7\,\,\, & 11.63 &  9.73 &  7.53 \\
 10317.6 & 13.5\,\,\, & 11.49 &  9.74 &       \\
 10392.4 & 13.80 & 11.62 &  9.74 &  7.72 \\
 10476.3 & 13.6\,\,\, & 11.61 &  9.72 &  7.6\,\,\, \\
 10764.5 & 13.9\,\,\, & 11.79 &  9.94 &  7.74 \\
 10796.4 & 13.8\,\,\, & 11.79 &  9.89 &  7.8\,\,\, \\
 11097.5 & 13.76 & 11.71 &  9.79 &  7.63 \\
\multicolumn{5}{l}{IRAS\,04553--6825   (WOH\,G064)}\\
  9797.3 &  9.88 &  8.26 &  7.23 &  5.42 \\
  9960.6 &  9.83 &  8.25 &  7.24 &  5.45 \\
 10029.5 &  9.77 &  8.18 &  7.18 &  5.34 \\
 10144.2 &  9.64 &  8.06 &  7.05 &  5.18 \\
 10317.6 &  9.56 &  7.92 &  6.93 &  5.10 \\
 10446.4 &  9.60 &  7.98 &  6.99 &  5.19 \\
 10474.4 &  9.65 &  8.01 &  7.01 &  5.27 \\
 10761.5 &  9.80 &  8.22 &  7.25 &  5.48 \\
 10794.5 &  9.82 &  8.27 &  7.27 &  5.50 \\
 10917.3 &  9.69 &  8.12 &  7.11 &  5.24 \\
\multicolumn{5}{l}{IRAS\,04559--6931  (HV\,12501)}\\
  9705.3 &  8.70 &  7.88 &  7.67 &  7.32 \\
  9747.4 &  8.72 &  7.87 &  7.66 &  7.24 \\
  9959.7 &  8.71 &  7.89 &  7.63 &  7.25 \\
 10031.4 &  8.77 &  7.92 &  7.67 &  7.31 \\
 10120.3 &  8.88 &  8.04 &  7.77 &  7.36 \\
 10325.7 &  8.79 &  7.95 &  7.73 &  7.37 \\
 10392.5 &  8.78 &  7.95 &  7.69 &  7.43 \\
 10472.5 &  8.74 &  7.94 &  7.70 &  7.32 \\
\multicolumn{5}{l}{IRAS\,05042--6720  (HV\,888)}\\ 
  4941.5 &  7.86 &  7.03 &  6.78 &       \\
  5016.3 &  7.74 &  6.94 &  6.68 &  6.3\,\,\, \\
  5705.3 &  7.99 &  7.20 &  6.88 &   *   \\
  6066.5 &  7.99 &  7.10 &  6.78 &  6.36 \\
  6358.5 &  8.06 &  7.21 &  6.91 &    *   \\
  6425.3 &  8.07 &  7.22 &  6.92 &    *   \\
  6444.4 &  8.06 &  7.21 &  6.91 &    *   \\
  6489.4 &  8.04 &  7.21 &  6.89 &    *   \\
  6784.4 &  7.86 &  6.99 &  6.68 &    *   \\
  9705.3 &  7.97 &  7.06 &  6.78 &  6.38 \\
  9960.7 &  8.09 &  7.15 &  6.85 &  6.40 \\
 10031.6 &  8.10 &  7.19 &  6.88 &  6.44 \\
 10124.3 &  8.10 &  7.21 &  6.89 &  6.4\,\,\, \\
 10325.6 &  8.10 &  7.18 &  6.89 &  6.48 \\
 10476.4 &  8.00 &  7.08 &  6.79 &  6.37 \\
 11834.6 &  7.78 &  6.89 &  6.60 &  6.16 \\
\multicolumn{5}{l}{IRAS\,05128--6723   (HV\,2360)}\\
  4922.5 &  8.88 &  8.02 &  7.67 &   *   \\
  5677.4 &  8.85 &  7.97 &  7.64 &   *   \\
  6066.5 &  8.82 &  7.96 &  7.56 &  6.96 \\
  6357.5 &  8.83 &  7.94 &  7.57 &    *   \\
  6428.4 &  8.78 &  7.93 &  7.58 &    *   \\
  6490.4 &  8.77 &  7.90 &  7.57 &    *   \\
  6787.4 &  8.77 &  7.92 &  7.60 &    *   \\
  9705.4 &  8.88 &  8.02 &  7.66 &  7.15 \\
  9963.6 &  8.98 &  8.08 &  7.72 &  7.06 \\
 10028.4 &  9.00 &  8.11 &  7.76 &  7.14 \\
 10144.4 &  9.08 &  8.19 &  7.82 &  7.25 \\
 10325.6 &  8.87 &  8.02 &  7.70 &  7.12 \\
 10353.6 &  8.86 &  8.01 &  7.69 &  7.18 \\
 10390.6 &  8.84 &  8.01 &  7.67 &  7.12 \\
 10473.4 &  8.92 &  8.06 &  7.71 &  7.24 \\
\hline
\end{tabular}
\end{center}
\end{table}
\setcounter{table}{0}
\begin{table}
\begin{center}
\caption[]{continued}
\begin{tabular}{rrrrr}
\hline
  JD& $J$ & $H$ & $K$ & $L$ \\ 
--2440000& \multicolumn{4}{c}{(mag)}\\
\hline
\multicolumn{5}{l}{IRAS\,05148--6730  (HV\,916)}\\
  4920.6 &  8.84 &  7.91 &  7.57 &  *    \\
  5020.3 &  8.16 &  7.76 &  7.48 &  *    \\
  5680.4 &  8.71 &  7.80 &  7.47 &  *    \\
  6031.3 &  8.68 &  7.73 &  7.43 &   *    \\
  6113.3 &  8.68 &  7.76 &  7.46 &   *   \\
  6357.6 &  8.53 &  7.62 &  7.33 &   *   \\
  6427.4 &  8.62 &  7.67 &  7.38 &   *   \\
  6491.3 &  8.65 &  7.74 &  7.43 &   *   \\
  6798.5 &  8.78 &  7.81 &  7.49 &   *   \\
  9705.4 &  8.70 &  7.79 &  7.47 &  7.01 \\
  9961.6 &  8.58 &  7.65 &  7.36 &  6.93 \\
 10032.4 &  8.53 &  7.57 &  7.29 &  6.86 \\
 10122.4 &  8.41 &  7.51 &  7.22 &  6.77 \\
 10324.7 &  8.69 &  7.75 &  7.42 &  6.93 \\
 10447.5 &  8.75 &  7.84 &  7.53 &  7.03 \\
 10504.3 &  8.80 &  7.89 &  7.57 &  7.15 \\
 10762.6 &  8.71 &  7.79 &  7.51 &  7.06 \\
 10917.3 &  8.58 &  7.64 &  7.37 &  6.94 \\
\multicolumn{5}{l}{IRAS\,05316--6604}\\
  9965.5 & 10.15 &  9.28 &  8.78 &  7.81 \\
 10028.5 & 10.12 &  9.28 &  8.76 &  7.77 \\
 10142.4 & 10.29 &  9.47 &  8.95 &  8.0\,\,\, \\
 10323.6 & 10.60 &  9.77 &  9.16 &  8.2\,\,\, \\
 10353.6 & 10.68 &  9.82 &  9.19 &  8.2\,\,\, \\
 10442.5 & 10.42 &  9.56 &  8.98 &  8.0\,\,\, \\
 10475.4 & 10.28 &  9.44 &  8.91 &       \\
 10504.4 & 10.22 &  9.41 &  8.90 &  8.0\,\,\, \\
 10798.4 &  9.90 &  9.10 &  8.64 &  7.69 \\
 10887.4 & 10.13 &  9.36 &  8.87 &  7.90 \\
 10915.3 & 10.20 &  9.41 &  8.89 &  7.96 \\
 11210.5 & 10.57 &  9.70 &  9.14 &  8.2\,\,\, \\
\multicolumn{5}{l}{IRAS\,05327--6757  (HV\,996)}\\                                         
  6357.5 &  9.17 &  8.23 &  7.69 &   *   \\
  6428.5 &  9.06 &  8.14 &  7.67 &   *   \\
  6490.4 &  8.98 &  8.09 &  7.63 &   *   \\
  6801.5 &  8.84 &  7.94 &  7.51 &   *   \\
  9705.3 &  8.91 &  7.95 &  7.49 &  6.76 \\
  9746.6 &  8.97 &  8.00 &  7.51 &  6.76 \\
 10029.5 &  8.98 &  8.07 &  7.64 &  6.83 \\
 10123.3 &  8.94 &  8.04 &  7.59 &  6.83 \\
 10317.6 &  8.94 &  7.99 &  7.56 &  6.86 \\
 10447.5 &  8.99 &  8.05 &  7.62 &  6.91 \\
 10476.3 &  9.02 &  8.06 &  7.61 &  6.91 \\
\multicolumn{5}{l}{SHV\,05221--7025}\\
  9963.6 & 13.6\,\,\, & 11.99 & 10.83 &       \\
 10032.5 & 14.4\,\,\, & 12.63 & 11.17 &       \\
 10327.6 & 12.97 & 11.46 & 10.41 &       \\
 10499.4 & 13.92 & 12.22 & 10.98 &       \\
 10704.6 & 13.4\,\,\, & 11.88 & 10.76 &       \\
 10796.5 & 14.4\,\,\, & 12.78 & 11.35 &       \\
 10889.3 & 14.5\,\,\, & 12.49 & 11.32 &       \\
 10917.2 & 14.6\,\,\, & 12.68 & 11.29 &       \\
\multicolumn{5}{l}{SHV\,05357--7024}\\
  9746.5 & 12.03 & 11.03 & 10.76 &       \\
 10029.6 & 12.08 & 11.01 & 10.76 &       \\
 10142.5 & 12.14 & 11.12 & 10.94 &       \\
 10201.3 & 12.06 & 10.99 & 10.76 &       \\
 10201.3 & 12.58 & 11.27 & 10.46 &       \\
 10323.7 & 13.26 & 11.70 & 10.75 &       \\
 10442.5 & 12.59 & 11.28 & 10.52 &       \\
 10473.4 & 12.58 & 11.21 & 10.50 &       \\
\hline
\end{tabular}
\end{center}
\end{table}
\setcounter{table}{0}
\begin{table}
\begin{center}
\begin{tabular}{rrrrr}
\hline
  JD& $J$ & $H$ & $K$ & $L$ \\ 
--2440000& \multicolumn{4}{c}{(mag)}\\
\hline
\multicolumn{5}{l}{SHV\,05357--7024 continued}\\
 10498.5 & 12.53 & 11.22 & 10.45 &       \\
 10887.4 & 12.88 & 11.48 & 10.62 &       \\
 10914.2 & 13.11 & 11.62 & 10.72 &       \\
 11890.5 & 13.84 & 12.13 & 11.05 &       \\
 11982.3 & 13.35 & 11.84 & 10.85 &       \\
\multicolumn{5}{l}{GRV\,0519430--670044}\\
  9797.3 & 12.94 & 11.59 & 10.79 &       \\
  9961.6 & 12.10 & 11.01 & 10.58 &       \\
 10027.6 & 12.79 & 11.57 & 10.81 &       \\
 10144.4 & 13.11 & 11.71 & 10.89 &       \\
 10324.6 & 12.63 & 11.35 & 10.67 &       \\
 10442.5 & 12.93 & 11.63 & 10.86 &       \\
 10474.5 & 12.97 & 11.64 & 10.91 &       \\
 10504.3 & 12.83 & 11.61 & 10.93 &       \\
 10767.5 & 12.96 & 11.66 & 10.91 &       \\
 10797.4 & 12.85 & 11.61 & 10.94 &       \\
 10887.3 & 12.23 & 11.06 & 10.61 &       \\
 10915.3 & 12.28 & 11.08 & 10.57 &       \\
 11890.6 & 12.74 & 11.36 & 10.62 &       \\
 11949.4 & 13.05 & 11.71 & 10.85 &       \\
\multicolumn{5}{l}{GRV\,0519486--645415}\\
 10387.6 & 12.98 & 11.80 & 11.20 &       \\
 10442.4 & 12.59 & 11.41 & 10.97 &       \\
 10470.5 & 12.87 & 11.64 & 11.03 &       \\
 10498.3 & 13.29 & 11.98 & 11.26 &       \\
 10602.2 & 13.18 & 11.99 & 11.33 &       \\
 10701.6 & 12.65 & 11.48 & 10.97 &       \\
 10761.5 & 13.2\,\,\, & 12.00 & 11.24 &       \\
\multicolumn{5}{l}{GRV\,0530506--643714}\\
 10443.4 & 13.5\,\,\, & 12.78 & 12.60 &       \\
 10471.4 & 12.9\,\,\, & 12.06 & 11.81 &       \\
 10498.4 & 12.72 & 11.85 & 11.54 &       \\
 10602.2 & 12.94 & 12.14 & 11.87 &       \\
 10706.6 & 13.5\,\,\, & 12.45 & 11.95 &       \\
 10762.6 & 13.11 & 12.46 & 12.5\,\,\, &       \\
\multicolumn{5}{l}{GRV\,0537140--674024}\\
 10387.6 & 12.98 & 11.47 & 10.45 &  9.6\,\,\, \\
 10442.4 & 12.09 & 10.75 & 10.07 &       \\
 10471.5 & 12.76 & 11.33 & 10.39 &       \\
 10498.4 & 12.05 & 10.70 & 10.01 &       \\
 10760.6 & 11.96 & 10.56 &  9.89 &       \\
 10762.5 & 11.96 & 10.55 &  9.88 &  8.8\,\,\, \\
\multicolumn{5}{l}{IRAS\,05289--6617}\\
 11181.5 & 14.2\,\,\, & 13.22 & 12.70 &       \\
\hline
\end{tabular}
\end{center}
\end{table}
\section{Introduction}
 Stars at the top of the Asymptotic Giant Branch (AGB) are of interest,
primarily for two reasons. First, they are undergoing mass loss --
via mechanisms which are still poorly understood. Secondly, they
represent the most luminous phase in the evolution of low- and
intermediate-mass stars and, as high resolution techniques enable us to
examine stars in ever more distant galaxies, are therefore among the first
individual objects to be isolated in a given population; furthermore, they
will contribute a significant fraction of the infrared luminosity of distant
unresolved populations (Ferraro et al. 1995; Mouhcine \&Lan\c{c}on 2002). The
Large Magellanic Cloud (LMC) provides a particularly useful laboratory for
the studies of luminous AGB stars at a known distance but with a range of
initial masses.

 Two factors complicate the studies of these luminous AGB stars - their long
period large amplitude variability, which necessitates observations over
periods of many years, and their high mass-loss rates which lead to thick
dust shells and necessitate observations at mid-infrared wavelengths.
Earlier papers in this series (Loup et al. (1997 - Paper~I), Zijlstra et al.
(1996 - Paper~II), van Loon et al. (1997, 1998, Papers~III and IV), Trams et
al. (1999b - Paper~V) and van Loon et al.\ (1999 - Paper~VI)) described the
identification of luminous AGB stars in the LMC, using IRAS and $JHKL$
photometry, as well as extensive ISO observations of a subset of them, and
the derivation of their mass-loss rates.
 
This paper concentrates on the variability characteristics of the sample,
reporting near-infrared ($JHKL$) photometry obtained in support of the
spectroscopy and mid-infrared photometry discussed in previous Papers (I to
VI). These data, together with others from the literature, are used to
determine pulsation periods and amplitudes and to look for long-term trends
in the behaviour of the stars. They are also combined with the observations
discussed previously to determine the bolometric magnitudes and examine
their relationship with the pulsation characteristics of the stars.

It is known that stars near to the end of their AGB evolution are large
amplitude, reasonably regular pulsators and that stars with these
characteristics form a well defined group (e.g. Whitelock 2002 and
references therein).  As the intention here is to consider stars very near to
the tip of the AGB, we use the information at our disposal to isolate the
objects of interest, removing from the sample supergiants as well as
semi-regular and irregular pulsators.

Following Feast, Whitelock \& Menzies (2002) we assume here that the
distance modulus of the LMC is $(m-M)_0 = 18.60$ mag, and, where necessary,
the results of other analyses are corrected to this value.

\subsection{Hot Bottom Burning (HBB)}
  Towards the end of AGB evolution, in stars with initial masses in excess
of 4 or 5 $M_{\odot}$ (or perhaps at even lower masses for very low
metallicities), the base of the H-rich convective-envelope can dip into the
H-burning shell; the introduction of fresh H-rich material into the
nuclear-burning shell allows the luminosity to go above the core-mass
luminosity predictions (Bl\"ocker \& Sch\"onberner 1991). This process is
known as envelope burning or hot bottom burning. Carbon is burned to
nitrogen affecting the transition from oxygen-rich to carbon-rich,
although the details depend on the model and particularly on how mass loss
is treated. HBB may prevent C stars from forming at all until the envelope
mass is depleted (Frost et al. 1998), although in some models C stars do
form and HBB turns the stars back from C-rich to O-rich. Another
consequence, for stars in a rather narrow mass range, is the formation of
lithium and its mixing to the surface via the beryllium transport
mechanism (Sackmann \& Boothroyd 1992). Smith et al. (1995) surveyed
luminous AGB stars in the LMC and SMC for lithium - the clearest indication
that HBB is taking place.

Two stars in our sample are known to be undergoing HBB from the presence of
lithium in their spectra (Smith et al. 1995), viz, the O-rich star 
HV\,12070 and C-rich star
SHV\,05210--6904 (SHV\,F4488). Trams et al. (1999a) showed that the
luminous C star, IRAS\,04496--6958, has a silicate shell and suggested that it
has only very recently stopped HBB. These three stars are of particular
interest and are identified in the various diagrams and the discussion which
follows.

There are very few LMC stars whose spectra have been examined for lithium, 
so none of the others are actually known for certain {\it not} to be
experiencing HBB. Indeed we would guess that several other stars considered  
in this paper are in the same condition.

A preliminary analysis of the results from this paper, with an emphasis 
on the significance of the HBB stars in the period-luminosity relation,
was discussed by Whitelock \& Feast (2000). 

\section{Source Selection and $JHKL$ photometry}
  The coordinates and alternative names of the sources discussed here were
listed in Paper~I, while Paper~II gives the results of the first batch of
$JHKL$ photometry. In addition to the sources discussed in previous papers
in this series we include a few AGB variables previously observed from the
South African Astronomical Observatory (SAAO) for which new observations
have been obtained. These stars have only thin dust shells, but provide a
useful link with the work of Feast et al. (1989) who established a
period-luminosity (PL) relation from $JHK$ observations of these stars and
others like them.

Near-infrared ($JHKL$) photometry was carried out between 1993 and 2001
using the Mk~III photometer on the SAAO 1.9-m telescope at Sutherland and
transformed onto the SAAO standard system defined by Carter (1990).  Earlier
observations were obtained for some sources as part of other programmes and
some of the data published by Feast et al. (1980), Catchpole \& Feast (1981)
or Glass et al. (1990). The new results are listed in Table~\ref{table.JHKL}
which also gives the Julian Date (JD) of the observation. Measurements were
made through either 12 or 18 arcsec apertures depending on the seeing and
the crowding of the individual source. A few of the older observations were
made on the 0.75-m telescope through a 36 arcsec aperture -- they are marked
with an asterisk in the $L$ column. The magnitudes are accurate to better
than $\pm 0.05$ mag, unless they are quoted to one decimal place when they
are good to $\pm0.1$ mag. Many of the sources were too faint to measure at
$J$, $H$, or $L$, particularly near minimum light. It also remains possible
that some of the fainter magnitudes quoted at $JH$ have been slightly
contaminated due to the close proximity of other sources.

The $JHKL$ observation quoted in Paper~II for IRAS\,04498--6842 is actually
a measure for IRAS\,05003--6712. The observation for IRAS\,05289--6617 was
of the wrong source; only one observation was obtained for the correct
object.

The star RHV\,0524173--660913 was not part of the ISO programme and was
observed towards the end of our programme because it appeared to be a
particularly luminous C-star. The C spectral type comes from Reid, Hughes \&
Glass (1995) and its position in a PL relation is shown in fig.~1 of
Groenewegen \& Whitelock (1996) where it lies well above the standard
relation.  However, we identify it here with the star that Westerlund,
Olander \& Hedin (1981) classified as an M giant and which is known as
WOH\,G311 (or WOH\,G310). It is considered to be an O-rich star in the
following discussion, but another spectrum is required to settle the matter. 
If it really is an M star that has turned into a C star it is obviously
extremely interesting and perhaps similar to IRAS\,04496--6958 (Trams et al.
1999a).

%
%
\setcounter{table}{1}
\begin{table*}
\begin{center}
\caption[]{\label{table.mean} Mean magnitudes, amplitudes and 
periods for the LMC sources}
\begin{tabular}{lrrrrrllllrrl}
\hline
  name& $J$ & $H$ & $K$ & $L$ & n & $\Delta J$ & $\Delta H$ & $\Delta K$ &
$\Delta L$ &  $P_K$ & $P_{other}$ & note\\
\hline
\multicolumn{12}{l}{\it O-rich AGB stars}\\
HV\,12070     & 10.43 &  9.48 &  9.04 &  8.4  & 21& 0.88& 1.02& 0.88& 0.8 &  623& 621&3,HBB \\
HV\,2446      & 10.49 &  9.50 &  9.06 &  8.4  & 16& 0.73& 0.87& 0.77& 0.6 &  596& 600&4 \\
IRAS\,04407--7000  & 11.13 &  9.67 &  8.79 &  7.62 & 27& 1.67& 1.53& 1.23& 1.08& 1199&      \\
IRAS\,04498--6842  &  9.94 &  8.87 &  8.08 &  7.16 & 20& 1.82& 1.66& 1.30& 1.23& 1292&      \\
IRAS\,04509--6922  & 10.80 &  9.42 &  8.59 &  7.56 & 25& 2.01& 1.82& 1.45& 1.19& 1292& 1290&1 \\
IRAS\,04516--6902  & 11.04 &  9.55 &  8.72 &  7.67 & 21& 1.76& 1.71& 1.41& 1.38& 1091&1090&1 \\
IRAS\,04545--7000  &       & 12.57 & 10.13 &  7.84 & 28&     & 1.81& 1.57& 1.45&*1216&1270&1 \\
IRAS\,05003--6712  & 12.90 & 11.26 &  9.95 &  8.49 & 12& 1.97& 1.92& 1.59& 1.44&  883&      \\
IRAS\,05294--7104  & 12.18 & 10.33 &  9.21 &  7.94 & 16& 1.8 & 1.5 & 1.2 & 1.2 & 1079&1040&1 \\
IRAS\,05329--6708  &       & 12.7\,\,\, &  9.9  &  7.6  & 43&     & 2.2 & 1.5 & 1.5 & 1262& 1295&2,W \\
IRAS\,05402--6956  &       &       & 10.4  &  8.0  & 21&     &     & 1.8 & 1.4 & 1393&1390&1  \\
IRAS\,05558--7000  & 12.52 & 10.50 &  9.25 &  7.82 & 16& 1.85& 1.71& 1.42& 1.32& 1220&      \\
SHV\,04544--6848& 10.72 &  9.69 &  9.13 &  8.36 &  7& 1.21& 1.18& 0.97& 0.8 &  645&  728 &5 \\
SHV\,05220--7012& 13.02 & 12.06 & 11.73 &       &  7& 0.85& 0.9 & 0.74&     &  219&  204 &5\\
SHV\,05249--6945& 12.14 & 10.92 & 10.39 &       &  6& 1.1 & 0.9 & 0.8 &     &  425&  413 &5\\
SHV\,05305--7022& 11.96 & 11.00 & 10.57 &       &  8& 0.60& 0.75& 0.68&     &  362&  348 &5\\
R105        & 11.75 & 10.74 & 10.33 &       & 20& 0.70& 0.70& 0.63&     &  413& 420&6\\
WBP\,74      & 12.83 & 11.87 & 11.50 &       & 16& 0.53& 0.51& 0.48&     &  245&  227&8 \\
GRV\,0517584--655140& 13.39 & 12.56 & 12.25 &       & 15& 0.58& 0.54& 0.46&     &  116&117&6     \\
RHV\,0524173--660913& 11.04 & 10.09 &  9.69 &       & 10& 0.79& 0.89& 0.81&     &  490&486&5 \\
\multicolumn{12}{l}{\it C-rich AGB stars}\\
IRAS\,04286--6937  & 15.0\,\,\, & 12.77 & 10.95 &  8.82 & 13&     & 1.28& 1.13& 1.03&  662&      \\
IRAS\,04374--6831  &       & 14.20 & 11.96 &  9.26 &  9&     &     & 1.44&     &  639&      \\
IRAS\,04496--6958  & 12.57 & 10.63 &  9.18 &  7.62 & 19& 1.07& 1.00& 0.88& 0.67& 723&& HBB \\
IRAS\,04539--6821  &       &       & 12.51 &       &  9&     &     & 1.65&     &  676&   \\
IRAS\,04557--6753  &       & 13.93 & 11.78 &  9.11 & 16&     & 1.42& 1.36& 0.93&  765&      \\
IRAS\,05009--6616  &       & 13.06 & 11.29 &  8.94 & 23&     & 1.5 & 1.5 & 1.4 & *658 & \\
IRAS\,05112--6755  &       & 15.1\,\,\, & 12.39 &  9.13 & 27&     &     & 1.75& 1.35& 830&822&2,W\\
IRAS\,05113--6739  &       & 14.6\,\,\, & 12.39 &  9.1  & 27&     &     & 1.82&     & 700&713&2,W\\
IRAS\,05128--6455  & 14.1\,\,\, & 12.46 & 10.91 &  9.0  & 17& 1.3 & 1.28& 1.06& 1.0 &  708&      \\
IRAS\,05190--6748  &       &       & 12.82 &       & 29&     &     & 1.66&     &
939&889&2\\
IRAS\,05291--6700  & 13.04 & 11.33 & 10.17 &  8.86 &  8& 0.87& 0.74& 0.52& 0.24&  483&      \\
IRAS\,05295--7121  &       & 12.96 & 10.99 &  8.83 & 14&     & 1.28& 1.15& 0.92&  682&      \\
IRAS\,05300--6651  &       & 14.8  & 12.30 &       & 27&     &     & 1.62&     &*708&683& 2,W \\
IRAS\,05360--6648  &       &       & 12.91 &       & 27&     &     & 1.22&     &
538&530&2 \\
SHV\,05003--6817& 13.70 & 11.85 & 10.58 &  9.3  &  8& 1.4 & 1.0 & 0.77&     &  369&  396 &5\\
SHV\,05003--6829& 13.65 & 11.74 & 10.33 &  8.8  &  9& 1.26& 1.06& 0.88&     &  441&  477 &5\\
SHV\,05027--6924& 12.59 & 11.40 & 10.82 &       &  8& 0.80& 0.67& 0.42&     &  298&  310 &5\\
SHV\,05210--6904& 11.44 & 10.18 &  9.50 &  8.8  & 13& 1.07& 0.99& 0.65& 0.8 &  541& 524&5, HBB \\
SHV\,05260--7011& 13.6\,\,\, & 11.92 & 10.54 &       &  9& 0.8 & 0.89& 0.62&     &  373&436&5\\
WBP\,14      & 13.35 & 11.70 & 10.62 &       & 10& 0.79& 0.63& 0.42&     &  351&  325 &8\\
TRM\,72      &       & 13.1  & 11.2  &  8.9  & 25&     & 1.4 & 1.3 & 1.0 & *571& 631 & 2 \\
TRM\,88      & 13.4\ & 11.43 &  9.99 &  8.38 & 22& 1.0 & 0.93& 0.71& 0.42& *544& 565 & 2 \\
\multicolumn{12}{l}{\it Supergiants, semi-regular and irregular variables,
or insufficient data}\\
HV\,5870     &  9.41 &  8.32 &  7.99 &  7.48 & 20& 0.2 & 0.2 & 0.2 & 0.2 &320& 627? &3,4 \\
IRAS\,04530--6916  & 13.75 & 11.69 &  9.82 &  7.67 &  9& 0.2 & 0.2 & 0.2 & 0.3 &none&
1260:&1\\
IRAS\,04553--6825  & 9.69  & 8.09  & 7.09  & 5.28  &10 &0.32 &0.35 &0.34 &0.42
&841&930&1  \\
IRAS\,04559--6931  &  8.79 &  7.97 &  7.70 &  7.30 &  8& 0.2 & 0.1 & 0.1 & 0.1 &none & 675 \\
IRAS\,05042--6720  &  7.99 &  7.12 &  6.82 &  6.38 & 16& 0.2 & 0.2 & 0.2 & 0.2 &none &  850 \\
IRAS\,05128--6723  &  8.95 &  8.07 &  7.69 &  7.11 & 15& 0.3 & 0.3 & 0.3 & 0.3 &none & 409 \\
IRAS\,05148--6730  &  8.66 &  7.77 &  7.43 &  6.97 & 17& 0.28& 0.28& 0.24& 0.28&  951&  743 \\
IRAS\,05316--6604  & 10.29 &  9.46 &  8.92 &  7.94 & 12& 0.8 & 0.7 & 0.6 & 0.5 &none&\\
IRAS\,05327--6757  &  8.97 &  8.04 &  7.59 &  6.8  & 11& 0.2 & 0.2 & 0.2 & 0.1 &  595& 760 &5 \\
SHV\,05221--7025& 13.8\,\,\, & 12.11 & 10.93 &       &  8& 1.5 & 1.26& 0.82&     &  & 384:&5,S?\\
SHV\,05357--7024& 12.89 & 11.46 & 10.60 &       &  5&$>0.7$&$>0.5$&$>0.3$&&none&\\
GRV\,0519430--670044& 12.70 & 11.44 & 10.77 &       & 14& 0.82& 0.62& 0.30&     &  312& 314&7     \\
GRV\,0519486--645415& 12.87 & 11.74 & 11.13 &       & 17& 0.69& 0.61& 0.33&     &  236& 242&6 \\
GRV\,0530506--643714& 13.12 & 12.42 & 12.08 &       & 18& 1.3 & 1.2 & 1.1 &     &  157& 157&7\\
GRV\,0537140--674024&       &       &       &       & 15&     &     &     &     &  odd& 418 &6\\
IRAS\,05289--6617  & 14.20 & 13.22 & 12.70 &       &  1&     &     &     &     &     &      \\
\end{tabular}
\end{center}
{\bf Notes to Table \ref{table.mean}:} {\bf HBB}: these objects are
thought to be, or have recently been, undergoing HBB. {\bf W}: the mean $H$
mag was estimated using data from Wood et al. (1992) and/or Wood (1998). 
{\bf S?}: possibly S spectral type (Hughes \& Wood 1990).  
References for $P_{other}:$ 1 Wood et al. (1992), 2 Wood (1998), 3
Gaposchkin (1970), 4 Payne-Gaposchkin (1971), 5 Hughes \& Wood (1990), 6
Feast et al. (1989), 7 Reid, Glass \& Catchpole (1988), 8 Wood, Bessell \&
Paltoglou (1985).
\end{table*}

\begin{figure*}
\includegraphics[height=20.5cm]{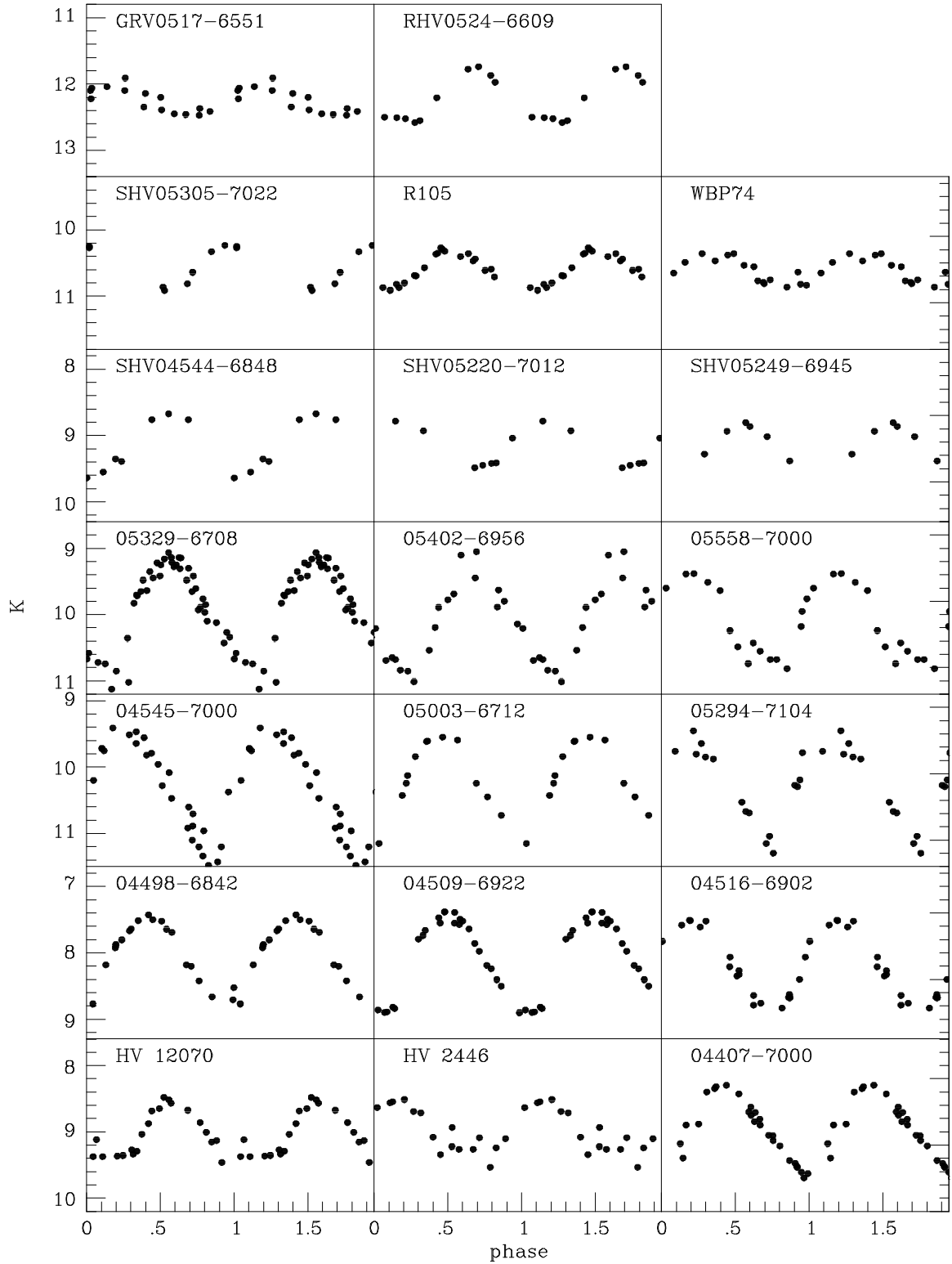}
\caption{\label{fig.lc1} $K$ light-curves for the O-rich AGB variables;
each point is plotted twice to emphasize the variability and the full height 
of each box is 2.6 mag. The phase is arbitrary (JD2\,440\,000 is phase zero
in all cases) and uses the period of $P_K$ from Table~\ref{table.mean}.}
\end{figure*}
\begin{figure*}
\includegraphics[height=23.4cm]{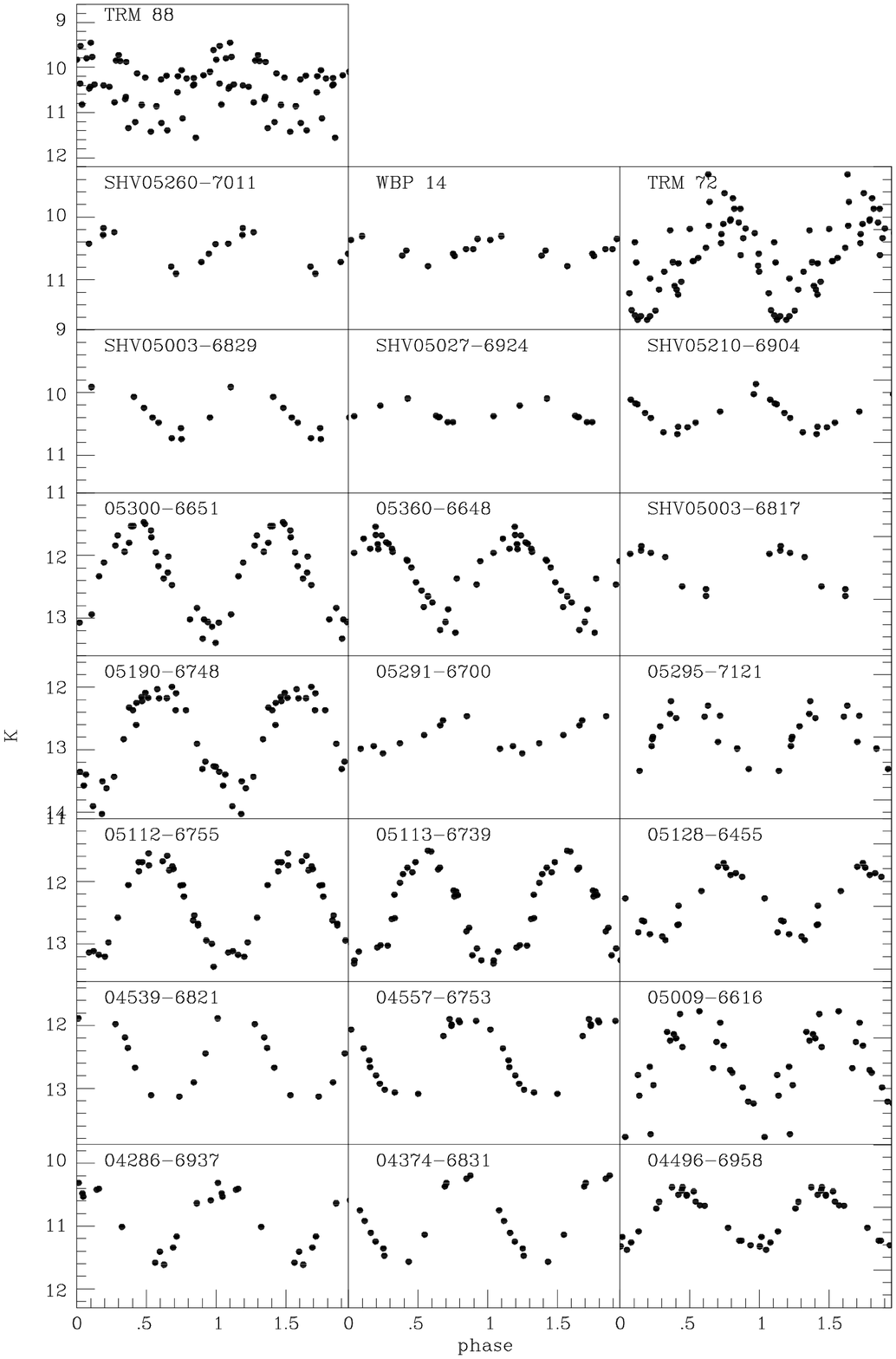}
\caption{\label{fig.lc2} $K$ light-curves for the C-rich AGB variables (see
Table~\ref{table.mean} for details).}
\end{figure*}

\section{Periods, Amplitudes and Mean Magnitudes}
 Table~\ref{table.mean} is divided into three parts, the first two
containing the summary data for the O- and C-rich AGB stars, using the
chemical types determined in earlier papers of this series or from Reid et
al. (1995). All of the sources in the first two parts of the table are
clearly large-amplitude variables with measurable periods. The Fourier-mean
magnitudes ($JHKL$) are listed, as are the number of $K$ observations (n)
used to determine the period ($P_K$) and the full peak-to-peak amplitude at
each wavelength ($\Delta J, \Delta H, \Delta K, \Delta L$). Periods from the
literature ($P_{other}$) are also listed and the appropriate references
given in the notes. There are 5 stars whose mean light is undergoing secular
or very long period variations, these are indicated with an asterisk in the
$P_K$ column and are discussed further in section 4.

The third part of Table~\ref{table.mean} contains all of the sources which
are not large amplitude AGB variables or have insufficient data to determine
periodicities. Seven of these are supergiants; they have $K<8.0$ mag, low
amplitude variability ($\Delta K < 0.4$ mag) and will have bolometric
magnitudes brighter than the tip of the AGB, i.e. $M_{bol}<-7.2$ mag. The
other 10 stars are semi-regular or irregular variables or stars without
sufficient data to classify them properly (it is possible that the
measurements of some of these stars were contaminated by the proximity of
other IR sources). The following discussion is limited to regular 
AGB-variables, i.e. leaving out the supergiants etc.

 The pulsation periods ($P_K$), listed in Table~\ref{table.mean}, were
determined from Fourier transforms of the $K$ light curves. Data from Glass
et al. (1990) was used in addition to that given in Table~\ref{table.JHKL},
while for stars in common with Wood et al. (1992) and/or Wood (1998) the
combined $K$ data sets, which provide very well sampled light curves, were
used for period determination. The results are generally in good agreement
with values determined by others ($P_{other}$) except for the semi-regular
and irregular variables where the agreement is often rather poor. Among the
large amplitude variables, our data for SHV\,05260--7011 does not fit the
436 day period determined by Hughes et al. (1990).

In the case of one O- and four C-rich stars the light curves show very
long-period or secular changes in addition to the obvious periodic
variations from pulsation. The periodicity of these long-term trends is very
uncertain and would require many more years of measurement to characterize.
The long term trends are discussed further in section 4. The peak-to-peak
amplitudes of the best fitting sine curves, given in Table~\ref{table.mean},
refer only to the periodic pulsations and the full variation of those stars
with long-term trends can be much larger than the values tabulated. A few of
the very long period O-rich stars have rather asymmetric light curves and
the tabulated amplitudes of the best fitting sine curves therefore
represent lower limits to the true peak-to-peak variations.
 
The mean magnitudes are determined from the Fourier fitted sine
curves. In the case of stars with a long term trend the mean used is 
that of the brightest cycles. As these means are used to determine the
bolometric magnitudes, this procedure might lead to an overestimate 
of the total flux, depending on what is causing the long term changes.

Fourier mean intensities are brighter than the mean magnitudes quoted here
by up to 0.3 mag at $K$, although for most stars the differences are around
$\Delta K \sim 0.1$ mag. The difference in bolometric magnitudes estimated
via intensity means is much smaller (see below). 

Figure~\ref{fig.pdk} shows a range of amplitudes at a given period, with the
amplitudes of the longer period C stars tending to be larger than those of
their O-rich counterparts, possibly because the C-star temperatures are
lower and therefore a steeper part of the blackbody light-curve is sampled
at these near-infrared wavelengths. The bolometric amplitudes may show much
less spread than do these $\Delta K$ values, but our information on them is
very limited. At a given period the stars with the larger $K$ amplitudes do
tend to have redder colours and higher mass-loss rates, as would be expected
if pulsation is driving the mass loss; compare, e.g. the two C stars
IRAS\,05113--6739 and IRAS\,05128-6455, both of which have periods around
700 days, but pulsation amplitudes of $\Delta K\sim$ 1.8 and 1.1 mag,
colours $K-[12]\sim$ 7.3 and 5.7 and mass-loss rates $\dot{M}\sim$
$1.1\times 10^{-5}$ and $0.6\times 10^{-5}$ $M_{\odot}\,\rm yr^{-1}$,
respectively (see Table~3). In general the amplitudes of these AGB stars are
comparable to those found by Olivier, Whitelock \& Marang (2001) among
galactic O- and C-rich stars with similar periods. The HBB stars all have
rather low amplitudes for their period, although not outside the range of
the others.

In a plot of $K-[12]$ against period or amplitude (see, e.g.
Fig.~\ref{fig.pk12} and Fig.~\ref{fig.dkk12}) there is clear separation of
the O- and C-rich stars, although it is interesting to note that the C-rich
HBB stars fall with O-rich stars rather than with the other C stars. The
separation between C- and O-types is also clear in the near-infrared colours
as illustrated in Fig.~\ref{fig.pkh} and Fig.~\ref{fig.pkl}.

\begin{figure}
\includegraphics[width=8.3cm]{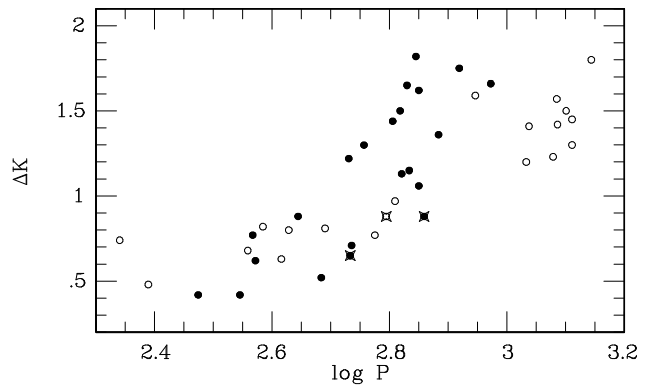}
\caption{\label{fig.pdk} The pulsation amplitude at $K$, $\Delta K$,
as a function of the period, $P$; solid and open symbols are C- and O-rich
stars respectively while HBB sources are starred.}
\end{figure}

\begin{figure} 
\includegraphics[width=8.3cm]{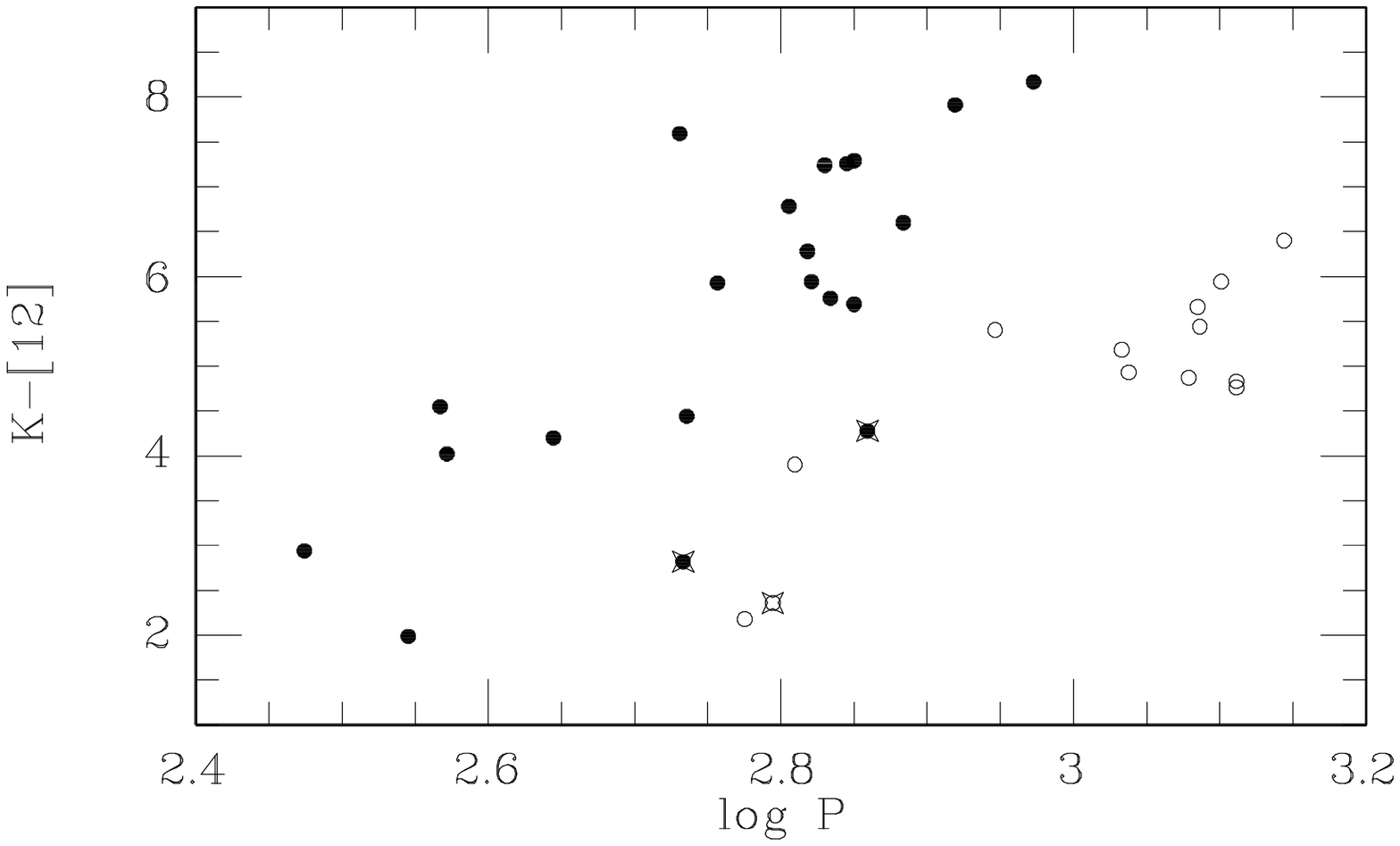} 
\caption{\label{fig.pk12} The $K-[12]$ colour as a function of the period, 
$P$; solid and open symbols are C- and O-rich stars respectively while  
HBB sources are starred.}
\end{figure}

\begin{figure}
\includegraphics[width=8.3cm]{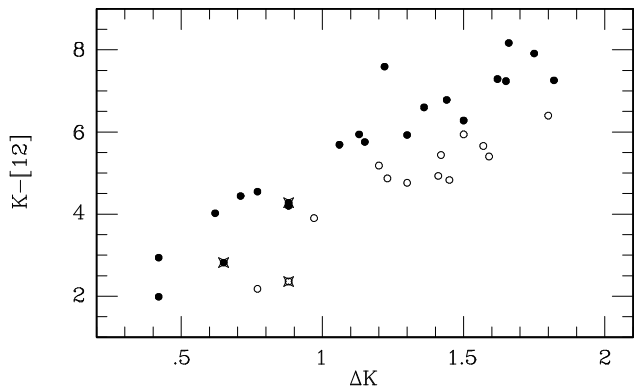}
\caption{\label{fig.dkk12} The $K-[12]$ colour as a function of the 
pulsation amplitude at $K$, $\Delta K$; solid and open symbols are C- and
O-rich stars respectively while HBB sources are starred.}
\end{figure}

\begin{figure} 
\includegraphics[width=8.3cm]{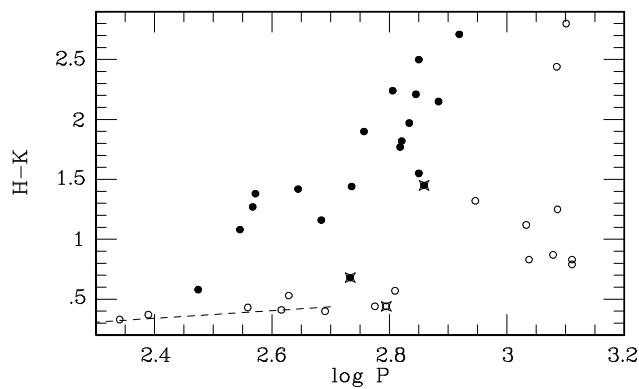} 
\caption{\label{fig.pkh} The $(H-K)$ colour as a function of the period, 
$P$; solid and open symbols are C- and O-rich stars respectively while
HBB sources are starred. The line is the period-colour relation for O-rich 
Miras in the LMC with $P<420$ days, from Feast et al. (1989).}
\end{figure}

\begin{figure} 
\includegraphics[width=8.3cm]{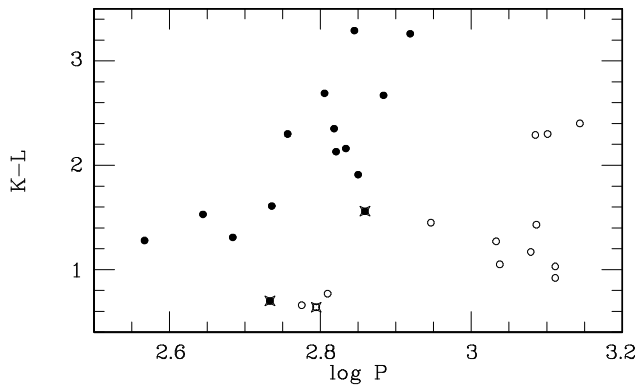} 
\caption{\label{fig.pkl} The $(K-L)$ colour as a function of the period, 
$P$; solid and open symbols are C- and O-rich stars respectively while
HBB sources are starred.}
\end{figure}
\section{long term trends}
\begin{figure}
\includegraphics[width=8.3cm]{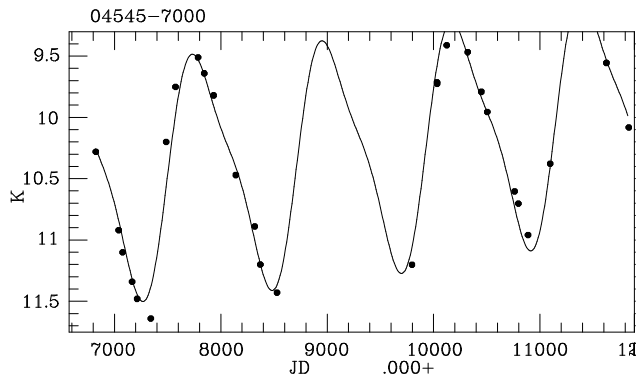}
\caption{\label{fig.04545} The $K$ light curve for IRAS\,04545--7000,
including data from Wood et al. (1992). The curve is a 1216 day sinusoid with
first harmonic and long term trend (illustrated using a 50\,000 day
sinusoid).}
\end{figure}

\begin{figure}
\includegraphics[width=8.3cm]{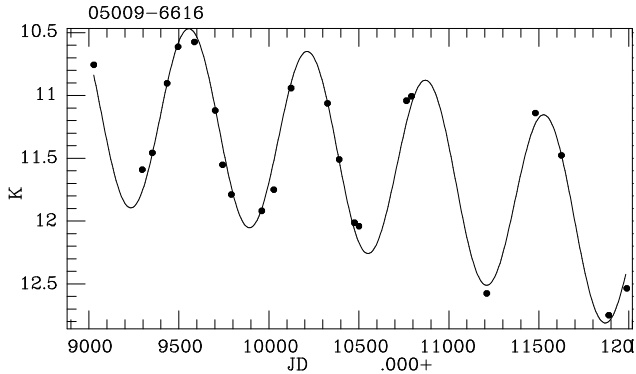}
\caption{\label{fig.05009} The $K$ light curve for IRAS\,05009--6616.
The curve is a 658 day sinusoid with a long term trend (illustrated using a
100\,000 day sinusoid).}
\end{figure}

\begin{figure}
\includegraphics[width=8.3cm]{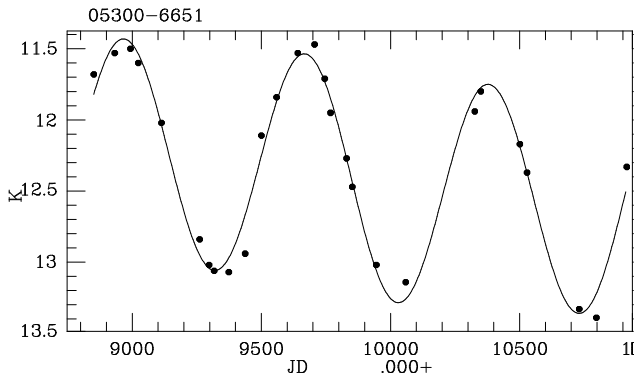}
\caption{\label{fig.05300} The $K$ light curve for IRAS\,05300--6651,
including data from Wood (1998). The curve is a 708 day sinusoid with a long
term trend (illustrated using a 2832 day sinusoid).}
\end{figure}

\begin{figure} 
\includegraphics[width=8.3cm]{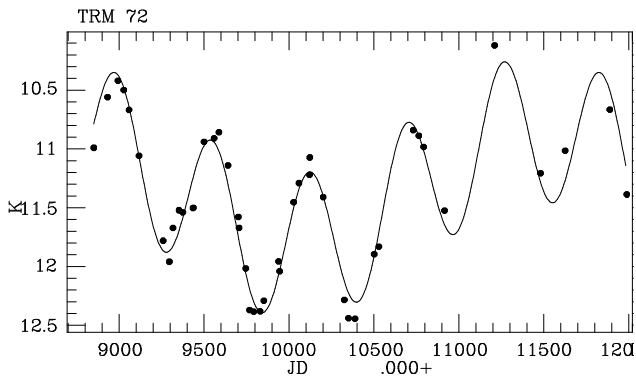}
\caption{\label{fig.trm72} The $K$ light curve for TRM 72, including data
from Wood (1998). The curve is a 571 day sinusoid with a long term trend
(illustrated using a 2855 day sinusoid).} 
\end{figure}

\begin{figure}
\includegraphics[width=8.3cm]{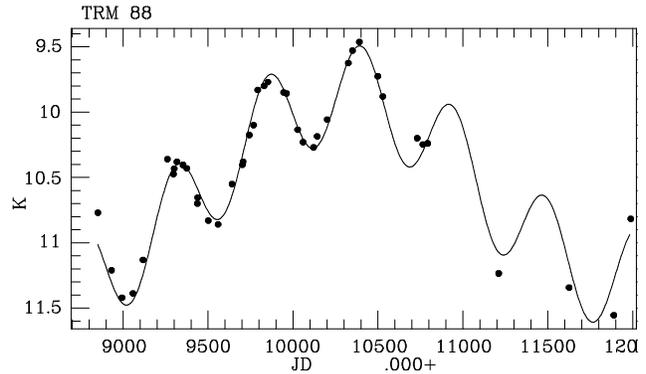}
\caption{\label{fig.trm88} The $K$ light curve for TRM 88, including data
from Wood (1998). The curve is 544 day sinusoid with a long term trend
(illustrated using a 3266 day sinusoid).} 
\end{figure}

 The $K$ light curves which show clear long term trends are illustrated in
Figs.~\ref{fig.04545} to \ref{fig.trm88}. The curves fitted to the data 
are described in the captions; all of them comprise two periods, the
main pulsation period as listed in Table~\ref{table.mean} and a
significantly longer period chosen to represent the long term trend.

For IRAS\,04545--7000, the only O-rich star which shows a trend, a harmonic
is also used to provide a much better fit to its distinctly asymmetric light
curve. For the four carbon stars a simple sinusoid provides a very good fit
to the main pulsational variation.

The time base is far too short to establish the periodicity, if any, of
these long term trends. Indeed it is only for TRM 72 and TRM 88 that we have
any real indication that the trends might by cyclical, and in these two
stars periods of $5P_K$ and $6P_K$, respectively, were used to fit the
trends. However, these long periods are not well constrained and
significantly different periods could provide equally good fits to the
trends.

The amplitudes of the trends are large, ranging from at least 0.5 up to
about 1.0 mag at $K$, i.e. comparable to the main pulsation amplitude. The
data at other wavelengths are not so extensive, but they are adequate to
show that trend amplitudes are wavelength dependent in roughly the same way
as the pulsation amplitudes, i.e. greater at shorter wavelengths.
  
Long term trends and erratic or secular variations have been observed in
various long period variables and seem to be more prevalent among C stars
than O-rich stars. In the case of V~Hya the $\Delta K \sim 1.7$ mag
variations over a 6160 day cycle (e.g. Olivier et al. 2001 and references
therein), are best explained as the result of orbital variations in a binary
system. In several other galactic C stars, most notably R~For, the long term
changes are erratic and best interpreted as the result of mass-loss
variations (e.g. Whitelock et al. 1997), possibly akin to RCB variations
which are caused by the ejection of ``soot'' clouds in random directions.

Winters et al. (1994) have predicted periodic mass-loss variations on time
scales of a few times the pulsation period.  It seems likely that the
trends described here are caused by mass-loss variations, but only very long
term monitoring will characterize them sufficiently well to see if they
fit the Winters et al. model.

\section{IRAS fluxes}
 The IRAS fluxes for most of these sources are listed in table~3 of Paper~V
where a comparison is also made with the fluxes measured by ISO. We use only
the 12 and 25 $\mu$m flux here, noting the high uncertainty of measurements
at longer wavelengths. The fluxes were colour-corrected only for the purpose
of estimating the bolometric magnitudes. Thus for the various figures the
IRAS magnitudes were calculated from: $[12]=\log 28.3/F_{12}$ and $[25]=\log
6.73/F_{25}$, for comparison with other papers in this series, but we note
that this process differs from that used by Olivier et al. (2001), for
example, who colour-correct the IRAS data used throughout their discussion.

%
%
\begin{table}
\caption[]{Absolute bolometric magnitudes derived using IRAS data
(where available).}
\begin{center}
\begin{tabular}{lccr}
\hline
name & $M_{bol}$ & $K-[12]$ & \multicolumn{1}{c}{$\log(\dot{M})$}\\
     &   \multicolumn{2}{c}{(mag)}\\
\hline
{\it O-rich} \\
 HV\,12070       & --6.48 &   2.36 &  --6.30\\   
 HV\,2446        & --6.40 &   2.18 &  --5.82\\   
 IRAS\,04407--7000   & --7.11 &   4.87 &  --4.30\\   
 IRAS\,04498--6842   & --7.72 &   4.76 &  --4.59\\   
 IRAS\,04509--6922   & --7.28 &   4.83 &  --4.77\\   
 IRAS\,04516--6902   & --7.11 &   4.93 &  --4.48\\   
 IRAS\,04545--7000   & --6.56 &   5.66 &  --3.55\\   
 IRAS\,05003--6712   & --6.20 &   5.40 &  --4.54\\   
 IRAS\,05294--7104   & --6.79 &   5.18 &  --4.19\\   
 IRAS\,05329--6708   & --6.95 &   5.94 &  --3.74\\   
 IRAS\,05402--6956   & --6.77 &   6.40 &  --3.74\\   
 IRAS\,05558--7000   & --6.97 &   5.44 &  --4.30\\   
 SHV\,04544--6848 & --6.45 &  3.90  &       \\   
 SHV\,05220--7012 & --3.87 &        & $<-8.00$\\
 SHV\,05249--6945 & --4.97 &        & --6.70\\
 SHV\,05305--7022 & --4.97 &        & --6.40\\
 R105         & --5.16  &        &       \\
 WBP\,74       & --4.07  &        &       \\
 GRV\,0517584--655140 & --3.67 &        &       \\
 RHV\,0524173--660913 & --5.88 &        &       \\
{\it C-rich}\\                      
 IRAS\,04286--6937   & --5.66 &   5.94 & --5.30\\   
 IRAS\,04374--6831   & --5.25 &   6.78 & --5.10\\   
 IRAS\,04496--6958   & --6.42 &   4.28 & --5.25\\   
 IRAS\,04539--6821   & --5.13 &   7.24 & --5.00\\   
 IRAS\,04557--6753   & --5.35 &   6.60 & --5.10\\   
 IRAS\,05009--6616   & --5.50 &   6.28 & --5.10\\   
 IRAS\,05112--6755   & --5.92 &   7.91 & --4.85\\   
 IRAS\,05113--6739   & --5.34 &   7.26 & --4.96\\   
 IRAS\,05128--6455   & --5.58 &   5.69 & --5.22\\   
 IRAS\,05190--6748   & --5.69 &   8.17 & --4.72\\   
 IRAS\,05291--6700   & --5.16 &        &       \\
 IRAS\,05295--7121   & --5.35 &   5.76 & --5.22\\   
 IRAS\,05300--6651   & --5.41 &   7.29 & --4.96\\   
 IRAS\,05360--6648   & --4.96 &   7.59 & --5.05\\   
 SHV\,05003--6817 & --5.05 &  4.55  &--5.72\\   
 SHV\,05003--6829 & --5.15 &  4.20  &--5.80\\   
 SHV\,05027--6924 & --4.44 &  2.94  &$<-8.00$\\   
 SHV\,05210--6904 & --5.70 &  2.82  &--6.22\\ 
 SHV\,05260--7011 & --4.76 &  4.02  &--6.10\\   
 WBP\,14        & --4.19 &   1.99 & --6.30\\   
 TRM\,72        & --5.22 &   5.93 & --5.30\\   
 TRM\,88        & --5.56 &   4.44 & --5.52\\                                 
\hline
\end{tabular}
\end{center}
\normalsize  
\end{table}  
             
\section{Bolometric Magnitudes}
 Bolometric magnitudes, listed in Table 3, were calculated by integrating
under a spline curve fitted to the mean $J_{74},H,K,L,$ 12- and 25-$\mu$m
fluxes as a function of frequency, following the procedure described in
section 6 of Whitelock et al. (1994).

The $HKL$ values used are those listed in Table~\ref{table.mean}. The $J$ mag
was converted back to the 74 inch system for consistency with Feast et al.
(1989); in fact the $J$ flux is negligible for almost all of these stars so
the conversion makes no practical difference. The 12- and 25-$\mu$m IRAS
fluxes (see section 5) were colour-corrected using the 12- to 25-$\mu$m
flux ratio, as described in the explanatory supplement of the PSC.

In cases where there was no magnitude available for particular wave bands
the following procedures were adopted for estimating a value to use in the
integration (based on relationships deduced from sources measured in all the
relevant bands):\\
{\it For O-rich stars:} \\
For the one star, SHV\,05402--6956, with no $H$ magnitude we use equation (3) from 
Paper~IV to derive $(H-K)$ from $K-[12]$.\\
For the eight stars without $L$ measures, the expression:
$(K-L)=0.239+0.946(H-K)$, which was derived from a least squares fit to the
colours of the other O-rich stars, was used to derive $(K-L)$ from $(H-K)$. This
assumption is important only for SHV\,05221--7025 which is moderately red
with $(H-K)=1.18$ and thus $(K-L)=1.36$; the others all have $(K-L)<0.75$.\\
For the four stars without IRAS 12-$\mu$m fluxes and the three with upper limits only,
we use equation (3) from Paper~IV to derive $K-[12]$ from $(H-K)$. Stars
without 25-$\mu$m fluxes are assumed to have uncorrected $[12]-[25]=0.9$ mag.\\ 
{\it For the C-rich stars:}\\
For three stars without $H$ magnitudes we use equation (2) of Paper IV to derive $(H-K)$
from $K-[12]$.\\
For six stars without $L$ mags, $L$ is estimated from $[12]$ via 
$K-[12]=1.737+1.838(K-L)$ a relation derived from a least squares fit to
the other C stars. For WBP14, $(K-L)=0.4$ is assumed as its IRAS flux
is anomalously faint.\\
Only one star has no 12-$\mu$m flux, IRAS\,05291--6700, and this is estimated
from equation (2) of Paper IV. The three stars without 25-$\mu$m fluxes are 
assumed to have $[12]-[25]=0$.

In all cases without a $J$ measure, O- or C-rich, the relation
$(J-H)=1.4(H-K)$ is assumed to derive it.

Because it is only the stars which have most of their flux at long
wavelengths that lack $J$ or $H$ measures, and vice versa, most of these
estimates of the missing quantities have little effect on the derived
bolometric luminosity. Thus it is only the assumptions about the $L$
magnitudes
that are potentially important for the bolometric magnitude.

A comparison of Fourier mean magnitudes and intensities for the $JHKL$
observations, indicates that the effect of using intensity means rather than
magnitude means is to brighten the bolometric magnitude by between 0.01 and
0.06 mag. Using intensities would therefore have a negligible affect on the
discussion and conclusions of this paper. Bolometric amplitudes are
discussed briefly in section 7.1.

\subsection{Bolometric Correction}

\begin{figure} 
\includegraphics[width=8.3cm]{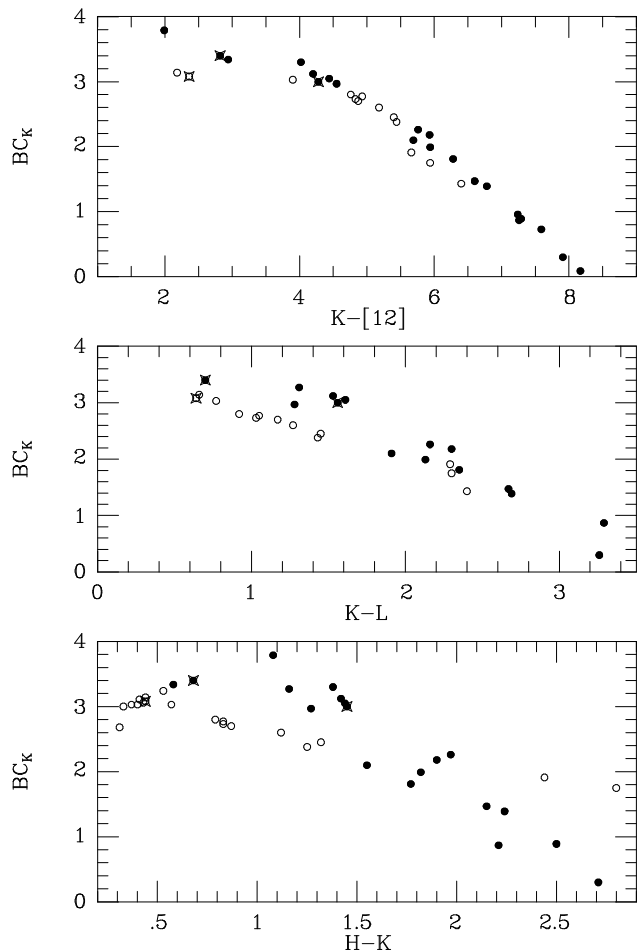} 
\caption{\label{fig.bc} The bolometric correction to $K$, BC$_K$, as a
function of various colours; solid and open symbols are C- and O-rich stars
respectively while HBB sources are starred.}
\end{figure}

 Figure \ref{fig.bc} shows the bolometric correction to $K$ as a function of
various colours. It is illustrated here because it is potentially useful for
application to other stars.  The $(K-L)$ and $(H-K)$ relationships are
distinctly different for O- and C-rich stars. For $K-[12]$, the relationship
is very similar for the two types, at least for the redder colours. Note
that, because the bolometric magnitude is determined by integrating under
the flux curve as a function of frequency, the bolometric correction is not
independent of the colours against which it is plotted.

\subsection{Comparison with Feast et al. (1989)}
 Feast et al. established PL relations from 29 O-rich and 20 C-rich LMC
Miras after deriving bolometric magnitudes by fitting blackbodies to 
fluxes from mean $JHK$ photometry -- all that was available at the time.  

Fitting blackbodies is an effective way of deriving bolometric magnitudes
if the dust shells make a negligible contribution to the total luminosities,
as they do for most short period Miras. It is obviously important to
look for any systematic differences in the blackbody and spline-fit methods
of establishing total luminosity. For this purpose we use both methods 
to estimate the total flux for sources with $P<441$ days and make a
comparison in Table 4.

The O-rich stars used by Feast et al. (1989) to establish the PL relation
all have $(J-K)<1.5$ mag. The six O-rich stars in Table 4 with $(J-K)<1.5$
mag have bolometric and spline luminosities that agree to better than 0.2
mag, five out of six agree to better than 0.04 mag. SHV\,05221--7025 shows a
slightly brighter $m_{bol}$ from the spline fit as would be expected for
stronger dust shell contribution; note, however, that SHV\,05221--7025
(which is classed among the SRs in Table~2) is the only source in this
sample that has an IRAS flux measurement whereas the others were all
estimated from the near-IR colours.

The C-rich stars used by Feast et al. (1989) to establish the PL relation
all have $(J-K)<3$ mag. Stars with similar colours in this sample have
spline- and blackbody-fit magnitudes that agree to better than 0.2 mag
(except for WBP\,14, see section 6 and Paper V fig~2 where the ratio of the
ISO to revised-IRAS flux for WBP\,14 is 2.2 -- the largest value found). The
redder stars tend to have slightly brighter spline-fit magnitudes. In this
case all but one of the stars have measured IRAS fluxes.

In summary, blackbody- and spline-fit bolometric magnitudes agree to better
than 0.2 mag (usually better than 0.1 mag) for O-rich stars with
$(J-K)<1.5$ mag and for C-rich stars with $(J-K)<3$ mag. For stars with larger
values of $(J-K)$ the blackbody method probably underestimates the total flux.
%
%
\begin{table}
\caption[]{A comparison of bolometric magnitudes derived from blackbody
(BB) and spline (spl) fits.}
\begin{center}
\begin{tabular}{lccrc}
\hline
name & $m_{bol}{\rm (BB)}$ & $m_{bol}{\rm(spl)}$ & BB-spl& $J-K$\\
\hline
{\it O-rich} \\
SHV\,05220--7012 &  14.71  &  14.73  &  --0.02 &  1.24 \\
SHV\,05221--7025 &  14.35  &  13.94  &   0.41 &  2.78\\
SHV\,05249--6945 &  13.64  &  13.63  &   0.01 &  1.69\\
SHV\,05305--7022 &  13.65  &  13.63  &   0.02 &  1.34\\
R105       &  13.41  &  13.44  &  --0.03 &  1.37\\
WBP\,74     &  14.52  &  14.53  &  --0.01 &  1.28\\
GRV\,0517--6551 &  15.12  &  14.93  &   0.19 &  1.10\\
RHV\,0524--6609 &  12.74  &  12.72  &   0.02 &  1.30\\
{\it C-rich}\\
SHV\,05003--6817 &   13.96 &   13.55 &    0.41 &  3.03\\
SHV\,05003--6829 &   13.64 &   13.45 &    0.19 &  3.23\\
SHV\,05027--6924 &   14.09 &   14.16 &  --0.07 &  1.71\\
SHV\,05210--6904 &   12.84 &   12.90 &  --0.06 &  1.88\\
SHV\,05260--7011 &   13.90 &   13.84 &    0.06 &  2.98\\
WBP\,14     &   14.06 &   14.41 &  --0.35 &  2.65\\
IRAS\,05291--6700 &   13.59 &   13.44 &    0.15 &  2.78\\
\hline
\end{tabular}
\end{center}
\normalsize
\end{table}

\section{PL Relation}

\begin{figure} 
\includegraphics[width=8.3cm]{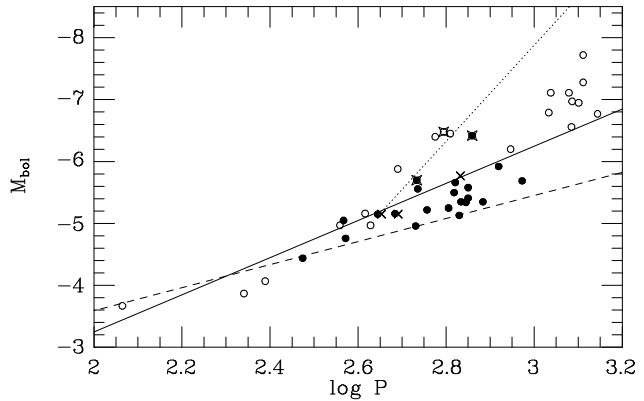} 
\caption{\label{fig.pl} The PL relationship for the LMC sources where the
bolometric mags have been calculated from IRAS data; solid and open
symbols are C- and O-rich stars respectively while HBB sources are starred and
the crosses represent C stars from LMC clusters (Nishida et al. 2000; van
Loon et al. 2003). The solid and dashed lines are extrapolations of the PL
relations derived by Feast et al. (1989) for O- and C-rich Miras,
respectively, with pulsations periods below 420 days. The dotted line is the
relation derived by Hughes \& Wood (1990) for AGB stars with periods over
400 days.}
\end{figure}

Figure~\ref{fig.pl} illustrates the PL relation ($M_{bol}$ calculated from
IRAS fluxes) for the stars under discussion and compares them with
extrapolations of the PL relations derived by Feast et al. (1989) for stars
with periods under 420 days. The PL relation derived by Groenewegen \& Whitelock
(1996) for C stars goes approximately through the middle of the two lines
from Feast et al. and would fit these C-star luminosities rather well.  WBP
14 is not illustrated in Fig.~\ref{fig.pl} as there are reasons to think its
IRAS flux is too faint. It is, however, illustrated in the IRAS/ISO
comparison made below (Fig.~\ref{fig.pl_comp}).

Nishida et al. (2000) discuss the luminosities and periods of three newly
discovered C-rich Miras in Magellanic Cloud clusters. The two from the LMC
clusters, NGC~1783 and 1978, are illustrated in Fig.~\ref{fig.pl}, where
they fall among the other C stars. van Loon et al. (2003) estimate the
period and luminosity for the carbon star, LI-LMC~1813, in the cluster
KMHK~1603 as (very approximately) 680 days and $M_{bol}=-5.77$.  The
turn-off masses for NGC~1783 and 1978 are around 1.5\,$M_{\odot}$ and that
of KMHK~1603 is around 2.2\,$M_{\odot}$. Taking this into account, and by
analogy with Olivier et al. (2001), we can deduce that the carbon stars in
this sample will mostly have had progenitors in the 1.5 to 2.5 $M_{\odot}$
range.

Feast et al. (1989) pointed out that stars with periods in excess of 420
days lay above the period luminosity relation, while Hughes \& Wood (1990) 
specifically derived a PL relation for stars with periods above 400 days
-- this is illustrated as a dotted line in the various PL diagrams shown
here.

\subsection{Comparison between bolometric magnitudes from ISO and IRAS data}

\begin{figure} 
\includegraphics[width=8.3cm]{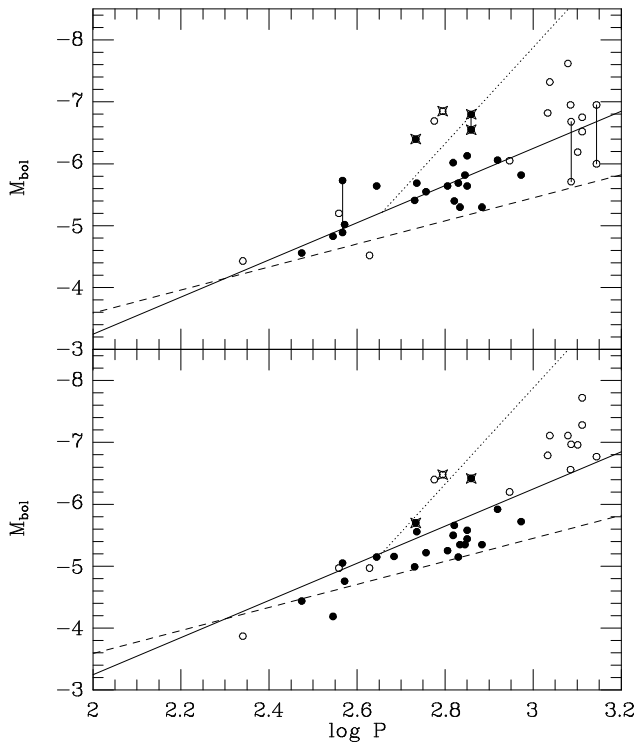} 
\caption{\label{fig.pl_comp} A comparison of the PL relations obtained using
IRAS (lower) and ISO (upper) fluxes, respectively, to calculate the
bolometric magnitude, $M_{bol}$. The IRAS $M_{bol}$ values are from Table~3
and the ISO ones from Paper~VI. The symbols and lines are described in the
caption to Fig.~\ref{fig.pl}.}
\end{figure}

\begin{figure} 
\includegraphics[width=8.3cm]{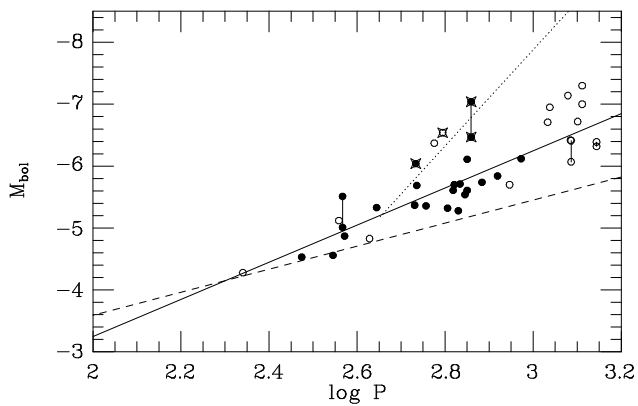}
\caption{\label{fig.pl_iso_cor} The ISO PL relation as shown in
Fig.~\ref{fig.pl_comp} (top),  after making phase correction to approximate 
the mean magnitude. The symbols and lines are described in the caption to 
Fig.~\ref{fig.pl}.}
\end{figure}

The differences between the IRAS and ISO fluxes for these LMC sources were
discussed in detail in Paper V, where the ISO fluxes were shown to be
systematically weaker than those measured by IRAS, although systematic
selection effects probably contribute to the differences. The fluxes for
these sources are close to the observational limit for the IRAS experiment,
while there is some doubt about the calibration of the ISO photometry. Thus
it is not clear which measurement should be regarded as more reliable.

A comparison of the long period ($P>1000$ days) O-rich sources in the two
parts of Fig.~\ref{fig.pl_comp} shows the bolometric magnitudes from IRAS
(${\bar M_{bol}}=-7.03$) to be systematically brighter than those from ISO
(${\bar M_{bol}}=-6.76$). In contrast the IRAS C-star magnitudes are clearly
fainter than those derived from ISO data. The difference in the behaviour of
the two types of star arises from the application of the colour-correction
to the IRAS data. The measured IRAS flux is divided by a factor, $f$, which
is a function of the ratio of the 12 to 25$\mu$m flux. This correction
changes the O-rich fluxes very little because the average value of $f$ is
1.01 and 1.22 for 12 and 25$\mu$m respectively, while for C stars the mean
$f$ is 1.19 and 1.31 for 12 and 25$\mu$m, respectively. It is not possible to
derive these corrections with any accuracy given that we have IRAS
fluxes only at 12 and 25$\mu$m for most of the sources, so this
correction procedure is far from ideal.

An interesting feature of the ISO observations relates to the four stars for
which measurements were obtained on different dates. The bolometric
magnitudes derived from these observations are shown in the upper part of
Fig.~\ref{fig.pl_comp} where observations of the same star are connected
with vertical bars. These illustrate that bolometric amplitudes of one
magnitude are quite possible (see also Papers IV and VI) and suggest that
some of the scatter seen in the various PL diagrams must be attributed to
variability. 

The greater scatter apparent in the upper (ISO) diagram in comparison to the
lower (IRAS) one can be understood in terms of the way the bolometric
magnitudes were derived for the two data sets.  For the ISO diagram they
were derived for the specific epoch at which the particular ISO observation
was obtained, while the IRAS diagram makes use of mean magnitudes, both for
$JHKL$ and from IRAS itself (although the IRAS mean is obviously not a true
value as IRAS observed for rather less than a year in total). 
Fig.~\ref{fig.pl_iso_cor} illustrates the PL relation derived using the ISO
data after making a correction to the mean magnitude (phases calculated from
dates given in Paper V). This correction was estimated by first calculating
the bolometric amplitude assuming $\Delta M_{bol}=\log P-2.0$, then assuming
that the bolometric variations are sinusoidal and determining the phase from
the $K$ light curve. While the corrections are obviously very uncertain, the
procedure does result in a very obvious reduction in the scatter over that
seen in Fig.~\ref{fig.pl_comp} (top).

If the process of deriving means for Fig.~\ref{fig.pl_iso_cor}
was perfect then all of the observations for a particular star would
coincide in this diagram, and that is obviously not the case.  In particular
the amplitudes of the HBB stars, which are probably low like their
$K$ amplitudes (see section 9) are overestimated as can be seen for
IRAS\,04496--6958.

It will be seen from the above discussion that it is not possible to make a
sensible estimate of the uncertainty associated with the bolometric
magnitudes. Variability is clearly significant and values derived from
single observations may differ from the mean by up to 0.5 mag, but there are
also systematic effects, possibly amounting to as much as a few tenths of a
magnitude, which depend on the way colour-corrections are treated.

\subsection{Comparison with Wood et al. (1992) and Wood (1998)}

\begin{figure} 
\includegraphics[width=8.3cm]{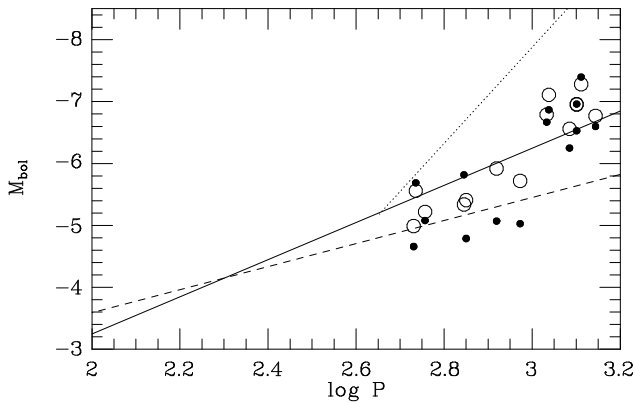} 
\caption{\label{fig.pl_wood} A comparison of the bolometric magnitudes
derived here (open circles) and those from Wood et al. (1992) and Wood
(1998) (closed circles) for the same sources - note that there are two
closed circles for IRAS\,05329--6708, one from each paper. The lines are
described in the caption to Fig.~\ref{fig.pl}.}
\end{figure}

Wood et al. (1992) obtained near-infrared photometry, measured periods and
determined bolometric magnitudes for six of the long period O-rich stars
discussed here, while Wood (1998) did the same for seven of the C stars and
one O-rich star (in fact that O-rich star, IRAS\,05329--6708, appears in both
papers).

The periods reported in these papers are very similar to ours; we determine
better values only by combining Wood's data with our own, as discussed in
section 3. However, the luminosities determined by Wood are distinctly
different from ours as can be seen from Fig.~\ref{fig.pl_wood}.  Both Wood
et al. (1992) and Wood (1998) derived bolometric magnitudes, as we do, from
a combination of near-infrared and IRAS fluxes, although the later paper also
applies an approximate correction to the IRAS flux to scale it to the mean
light. For the O-rich stars the mean from Wood's data for six stars,
$M_{bol}=-6.72\pm0.17$, compares to our value of $M_{bol}=-6.91\pm0.12$,
which is not significant, while for seven C stars the difference is larger
with Wood's data giving $M_{bol}=-5.14\pm0.17$ compared to our
$M_{bol}=-5.45\pm0.13$.  For IRAS\,05329--6708 there is a difference of 0.43
in the bolometric magnitudes quoted by Wood et al. (1992) and the fainter,
phase corrected, value quoted by Wood (1998), both points are illustrated in
Fig.~17.

Note that three of the C stars illustrated in Fig.~\ref{fig.pl_wood}
(IRAS\,05300--6651, TRM\,72, TRM\,88) are among those with long term trends
discussed in section 4. Nevertheless, the differences between our magnitudes
and those of Wood are not systematically influenced, in any significant way,
by our approach to establishing the mean near-infrared magnitudes for these
stars. In fact our bolometric magnitude ($-5.56$) for TRM\,88, which has
long term variations of around $\Delta K \sim 1.0$ mag, is slightly fainter
than that derived by Wood ($-5.69$).

\subsection{PL Relation - Summary}
 In conclusion we find that, omitting the stars undergoing HBB (see below),
both the C- and the O-rich variables tend to follow the trends suggested by
an extrapolation of the PL relation determined for shorter period ($P<420$
days) stars. There is no clear evidence that these sources with dust shells
deviate from the PL relation. They are, however, less luminous than HBB
stars with comparable periods. Owing to the uncertainty in the exact mean
values of the mid-infrared flux of these stars and therefore in their
bolometric magnitudes, it seems inappropriate to attempt to use these data
to better define the PL relation.

Wood (1998) particularly emphasizes that the luminosities of the dust
enshrouded stars (those with $500<P<900$ which we now know are C stars) fall
below the extrapolation of the short period PL relation. This is, however,
inconsistent with our findings discussed above.

\section{Mass-loss rates}
 Mass-loss rates were derived in Paper VI (and are listed in Table 3) for
many of the stars under discussion and these are used here to examine the
mass-loss dependence on the parameters discussed above. For
IRAS\,04496--6958 we follow Paper VI in adopting the mass-loss rate derived
by assuming the dust is pure carbon.
 
\begin{figure} 
\includegraphics[width=8.3cm]{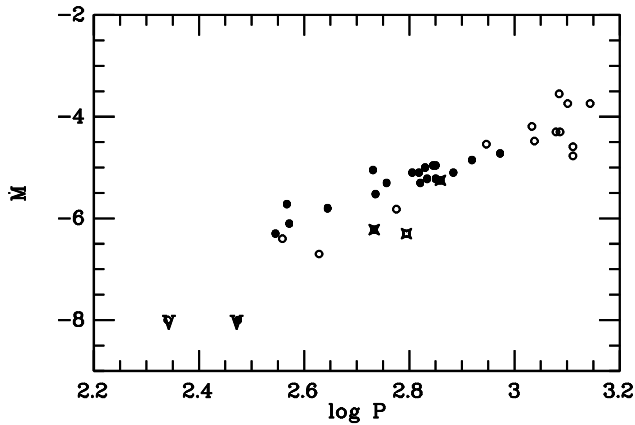} 
\caption{\label{fig.pdot} A plot of mass-loss rate, $\dot{M}$, against
pulsation period, $P$. Open and closed symbols are O- and C-rich stars
respectively, while stars thought to be undergoing HBB are starred and the
upper limits are marked with a V.}
\end{figure}

 The dependence of mass-loss rate on luminosity was discussed in Paper VI.
Given that these stars obey a PL relation we would anticipate a dependence
on period, as is indeed seen in Fig.~\ref{fig.pdot}. A very similar
dependence is seen for Galactic AGB variables (e.g. Whitelock et al. 1994;
Olivier et al. 2001). The relation is the same for O- and C-rich stars, just
as the PL relation derived from ISO data (Fig.~\ref{fig.pl_comp} (top) and
Fig.~\ref{fig.pl_iso_cor}) is similar for the two groups. The stars
suspected of HBB have lower mass-loss rates than other stars with similar
periods. This is of course well known, in that the $P>420$ day, luminous,
O-rich AGB stars, which we now think are undergoing HBB, were discovered
visually, while the C stars of comparable period have thick shells and were
only discovered via their infrared luminosity.

\begin{figure} 
\includegraphics[width=8.3cm]{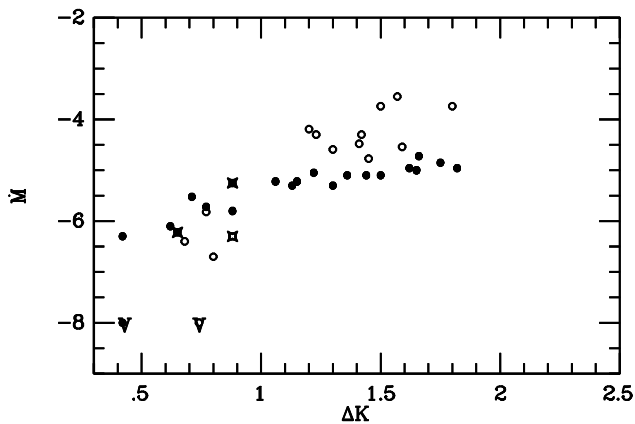}
\caption{\label{fig.dkdot} 
 A plot of mass-loss rate, $\dot{M}$, against pulsation amplitude, $\Delta
K$; symbols as for Fig.~\ref{fig.pdot}.}
\end{figure}

 Given that pulsation is thought to drive mass loss for these stars, some
dependence of mass-loss rates on amplitude is to be expected. However,
the pulsation amplitude and period are correlated (Fig.~\ref{fig.pdk}),
making the effects difficult to separate. Figure~\ref{fig.dkdot} shows
separate sequences for the O- and C-rich stars which cross around $\Delta K
\sim 1$ mag and $\log \dot{M} \sim -5.5$. At large amplitudes the O-rich
stars have both longer periods (higher luminosities) and higher mass-loss
rates than do their C-rich counterparts.  The HBB stars follow the same
trend as the others in this figure. 

\begin{figure} 
\includegraphics[width=8.3cm]{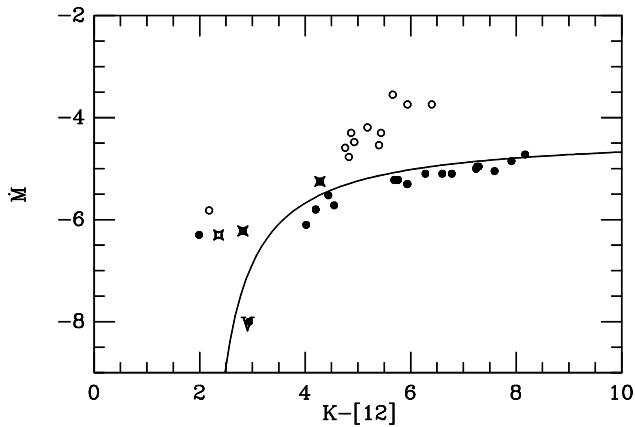} 
\caption{\label{fig.mdot} A plot of mass-loss rate, $\dot{M}$, against
$K-[12]$ colour; symbols as for Fig.~\ref{fig.pdot}. The curve is the
relation for galactic C stars derived by Olivier et al. (2001), to provide a
qualitative indication of the expectations at low mass-loss rates. }
\end{figure}

 $K-[12]$ is a measure of the optical depth of the dust shell
via the relative emission of the star
(which dominates the $K$ light) and the shell (which dominates [12]), and is
closely dependent on the mass-loss rate (Whitelock et al. 1994).
Figure~\ref{fig.mdot} shows rather different relationships for the O- and
C-rich stars in the LMC. Omitting for a moment the clump of four stars at
small $K-[12]$, most of the C- and O-rich stars follow
qualitatively similar relations to those found for high mass-loss AGB stars
in the solar neighbourhood (Olivier et al. 2001) although the O-rich stars
in the LMC reach higher mass-loss rates, $\dot{M}>10^{-4}\,M_{\odot}{\rm
yr^{-1}}$, at lower $K-[12]$ than do their Galactic counterparts. 
Galactic stars with $P>1000$ days have much redder colours as has been
discussed elsewhere (e.g. Wood et al. 1992).

The group of four stars with low $K-[12]$ and relatively high $\dot{M}$
comprises two of the three HBB stars, HV\,2446 which we would suspect is
also undergoing HBB and WBP\,14 which has too faint an IRAS flux and is
probably there erroneously. It is, perhaps, not surprising that the stars
undergoing HBB have a different mass-loss {\it vs} $K-[12]$ relation from
the other AGB stars.

\section{HBB Stars}
 We note that the light-curves for HV\,12070, which is known to be HBB, and
for HV\,2446 and RHV\,0524173--660913 (see Fig.~\ref{fig.lc1}),
which lie in a similar part of the PL diagram (Fig.~\ref{fig.pl_comp}), 
all have relatively low amplitudes (see Fig.~\ref{fig.pdk}). Furthermore,
they have similar shapes, exhibiting flat minima (the minima
of these particularly bright stars cannot be a consequence of field star 
contamination).

 HV\,12070, although O-rich, has an MS spectral type and a rather weak
10$\mu$m silicate feature (Paper V).  It is therefore likely that it is very
close to the O- and C-rich transition, although it is not possible to say in
which direction it is evolving.

 The three stars thought to be experiencing, or to have recently
experienced, HBB (see section 1.1) are more luminous than most sources of
comparable period. There are many more O-rich S-type stars that fall in the
same part of the PL diagram (Feast et al. 1989; Hughes \& Wood 1990), i.e.
close to the PL relation derived by Hughes \& Wood for stars with $P>400$
days. If the relatively high luminosities for these stars are a consequence
of HBB (see also Marigo, Girardi \& Bressan 1999) then they may also show
high lithium abundances and should be examined spectroscopically. It would
also be worth examining the O-rich stars with $P>1000$ days for lithium.

 It is interesting to note that there is a candidate HBB star in IC\,1613
which was discovered by Kurtev et al. (2001) and discussed by Whitelock
(2002) in the context of AGB stars in Local Group galaxies. It is O-rich,
with a spectral type of M3e, has a period of 641 days and is clearly more
luminous than an extrapolated PL relation. Its spectrum has not been
examined for lithium. Such stars are expected to be a feature of irregular 
galaxies with significant intermediate age populations.

\section{Conclusions}

Obscured AGB variables with $P>420$ days have luminosities close to the
extrapolation of the PL relation derived for stars with $P<420$ days. We
cannot rule out the possibility that there are fainter obscured AGB stars
which would not have been found to date because of the limited sensitivity
of infrared surveys of the LMC.

 We suggest that the apparent change in slope of the PL relation around
400-420 days, for stars with thin dust shells, is caused by the inclusion of
variables with luminosities brighter than the predictions of the core-mass
luminosity relation, owing to excess flux from HBB.

 If RHV\,0524173--660913 is confirmed to have changed from an M to a C
spectral type (see section 2) then it is of considerable interest.
It is presumably undergoing third dredge-up following a helium-shell flash,
possibly because hot bottom burning has just terminated.

\section*{Acknowledgments} We are grateful to Robin Catchpole, John
Menzies, Ian Glass and Brian Carter who made some of the $JHKL$ observations
published here.

\end{document}